\documentclass[aps,prb,reprint,twocolumn,showpacs,superscriptaddress]{revtex4-2}
\usepackage{graphicx}
\usepackage{pst-all}
\usepackage{color}
\usepackage{dcolumn}
\usepackage{amsmath}
\usepackage{amsthm}
\usepackage{bm}
\usepackage{diagbox}
\usepackage{layout}
\usepackage{float}
\usepackage{txfonts}
\usepackage{amsfonts}
\usepackage{amssymb}
\usepackage{array}
\usepackage{multirow,bigdelim}
\usepackage{enumitem}
\usepackage{ulem}

\definecolor{dukeblue}{rgb}{0.0, 0.0, 0.61}

\usepackage[colorlinks=true,linkcolor=dukeblue,pagecolor=dukeblue,
filecolor=dukeblue,menucolor=dukeblue,urlcolor=dukeblue,
citecolor=dukeblue,anchorcolor=dukeblue]{hyperref}

\begin{document}

\title{
Variational counterdiabatic driving of the Hubbard model for ground-state preparation
}

\author{Q.~Xie}
\affiliation{Quantum Computational Science Research Team, RIKEN Center for Quantum Computing (RQC), Saitama 351-0198, Japan}
\affiliation{Computational Condensed Matter Physics Laboratory, RIKEN Cluster for Pioneering Research (CPR), Saitama 351-0198, Japan}

\author{Kazuhiro~Seki}
\affiliation{Quantum Computational Science Research Team, RIKEN Center for Quantum Computing (RQC), Saitama 351-0198, Japan}

\author{Seiji~Yunoki}
\affiliation{Quantum Computational Science Research Team, RIKEN Center for Quantum Computing (RQC), Saitama 351-0198, Japan}
\affiliation{Computational Condensed Matter Physics Laboratory, RIKEN Cluster for Pioneering Research (CPR), Saitama 351-0198, Japan}
\affiliation{Computational Materials Science Research Team, RIKEN Center for Computational Science (R-CCS),  Hyogo 650-0047,  Japan}
\affiliation{Computational Quantum Matter Research Team, RIKEN Center for Emergent Matter Science (CEMS), Saitama 351-0198, Japan}

\date{\today}

\begin{abstract}

Counterdiabatic (CD) protocols
enable fast driving of quantum states
by invoking an auxiliary adiabatic gauge potential (AGP)
that suppresses transitions to excited states throughout the driving process.
Usually,  the full spectrum of the original unassisted Hamiltonian
is a prerequisite for constructing the exact AGP,
which implies that CD protocols are extremely difficult for many-body systems.
Here, we apply a variational
CD protocol recently proposed by
{P. W. Claeys \it{et al.}}
[\href{https://link.aps.org/doi/10.1103/PhysRevLett.123.090602}
{{Phys. Rev. Lett.} {\bf{123}}, 090602 (2019)}]
to a two-component fermionic Hubbard model in one spatial dimension.
This protocol engages an approximated AGP
expressed as a series of nested commutators.
We show that the optimal variational parameters in the approximated AGP
satisfy a set of linear equations
whose coefficients are
given by the squared Frobenius norms of these commutators.
We devise an exact algorithm that
escapes the formidable iterative matrix-vector multiplications
and
evaluates the nested commutators and the CD Hamiltonian in analytic representations.
We then examine the CD driving of the one-dimensional
Hubbard model up to $L=14$ sites with driving order $l\leqslant3$.
Our results demonstrate the usefulness of the variational CD protocol
to the Hubbard model and
permit a possible route towards
fast ground-state preparation for
many-body systems.

\end{abstract}

\maketitle


\section{Introduction}

Adiabatic control over quantum states
is of fundamental importance
for quantum information processing~\cite{NielsenMA2010},
quantum computation~\cite{RevModPhys.90.015002},
and many other dynamic processes~\cite{RevModPhys.71.S253}.
Nevertheless, adiabaticity can only be achieved
in sufficiently slow processes,
which inevitably expose the system
to dissipation and noise~\cite{JPhysSocJpn.5.435}.
This presents a major obstacle
and hinders many practical applications,
such as quantum state preparation~\cite{PhysRevA.105.032403}
and gate operations~\cite{PhysRevX.10.021054}.
Therefore, for the past few years,
there have been intensive efforts
aiming for protocols that speed up the evolution process and
at the same time render the desired system within the adiabatic regime.
These protocols are collectively called
shortcuts to adiabaticity (STA)~\cite{PhysRevLett.104.063002,
RevModPhys.91.045001}.

Some of the well-known STA are
the fast-forward~\cite{PhysRevA.78.062108,
ProcRSocA.466.1135},
adiabatic transfer~\cite{AnnuRevPhysChem.52.763,
OptLett.32.2771,
NatCommun.12.2156},
superadiabatic~\cite{NatPhys.13.330,
PhysRevA.95.012317,
SciAdv.5.eaau5999},
and stimulated Raman adiabatic passage protocols~\cite{JChemPhys.92.5363,
RevModPhys.70.1003,
PhysRevB.70.235317,
RevModPhys.89.015006}.
These protocols have found wide applications
in many fields of physics and chemistry,
including atomic, molecular, and optical physics,
condensed matter systems,
and quantum information and computation
(see Refs.~\cite{RevModPhys.70.1003,
RevModPhys.89.015006,
RevModPhys.91.045001} for reviews).
In particular,
applications to quantum computation
have drawn tremendous attention
due to recent rapid advances in quantum devices~\cite{Preskill2018quantumcomputingin,
Nature.574.505,
RevModPhys.94.015004},
as represented by the achievement of
intermediate-scale quantum chip packing more
than one hundred qubits~\cite{Nature.599.542}.
Indeed, to improve the fidelities of
quantum-state preparation and gate operations,
which are vital steps in quantum computing,
many STA have been proposed~\cite{PhysRevA.93.012311,
PhysRevA.100.012341,
PhysRevX.11.031070}.
Moreover, many STA protocols
have also been proposed
to boost the capabilities of quantum optimization algorithms~\cite{NewJPhys.21.043025,
PhysRevApplied.15.024038,
2021arXiv211208347H,
Wurtz2022counterdiabaticity,
PhysRevResearch.4.013141,
Sack2021quantumannealing,
2022arXiv220100790H}.

Counterdiabatic (CD) driving is one such powerful
STA protocol~\cite{TheJournalofPhysicalChemistryA.107.9937,
TheJournalofPhysicalChemistryB.109.6838,
JChemPhys.129.154111,
PhysRevLett.111.100502}.
The main idea behind the CD driving is to
add an auxiliary time-dependent term
to the original Hamiltonian
in such a way that
the states driven by
the resulting Hamiltonian evolve in time
along the trajectories of the instantaneous
eigenstates of the original one.
This additional term
is called the counter term and
takes the form~\cite{JPhysAMathTheor.42.365303}
\begin{align}
\label{eq:eq1}
\hat{\mathcal{C}}(t) =
- \text{i} \sum_{m,n\,(m\ne n)} \frac{|m\rangle \langle m| \left( \partial_t \hat{H} \right) |n\rangle \langle n|}{\epsilon_m - \epsilon_n},
\end{align}
where $\hat{H}$ is the original time-dependent Hamiltonian,
$|n\rangle$ is the
instantaneous eigenstate of $\hat{H}$ with its eigenenergy $\epsilon_n$,
and $\partial_t$ denotes the time ($t$) derivative.
The summation is performed over all eigenstates.
The counter term can exactly compensate
transitions to different eigenstates and
hence make the driving process transitionless~\cite{JPhysAMathTheor.42.365303}.
However, this expression immediately suggests its limitations
in two aspects:
(i) It is not well-defined when a level crossing occurs at, e.g., a phase transition,
as the spectrum gap closes at the crossing point.
(ii) It is difficult to implement
since it requires precise control of the
full spectrum over the driving period.

To circumvent these difficulties,
P. W. Claeys~{\it{et al.}}
have recently proposed
an approximated CD protocol~\cite{PhysRevLett.123.090602},
which adopts
a summation of
$l$ nested commutators
to mimic the exact counter term $\hat{\mathcal{C}}(t)$.
Here $l$ acts as an expansion order
and $l\to \infty$
retrieves the exact limit.
In such a strategy,
the full spectrum
is not necessary.
Another advantage is that
this protocol
shares a similar structure
with the Magnus expansion in periodically
driven systems.
Therefore, easy implementations
using Floquet engineering
can be expected,
although high driving frequencies
are required for large driving order $l$~\cite{PhysRevLett.123.090602}.
The approximated CD protocol
has been found useful for
spin systems, such as
the Ising model~\cite{PhysRevLett.123.090602}
and the $p$-spin model~\cite{PhysRevResearch.2.013283},
where ground-state fidelities
prepared by the approximated CD protocol
are increased with increasing driving order $l$
(Also see
Refs.~\cite{PhysRevLett.109.115703,
PhysRevA.90.060301,
JStatMechTheoryExp.2014.P12019,
PhysRevLett.114.177206,
PhysRevA.105.022618,
PhysRevE.87.062117,
JPhysSocJpn.86.094002,
PhysRevA.95.012309}
for more STA protocols
for quantum spin systems).
It is also applicable to
noninteracting fermion systems~\cite{ProcNatlAcadSci.114.E3909,
KOLODRUBETZ20171}.
Very recently,
a two-parameter generalization of this protocol
has been proposed and also demonstrated for the $p$-spin model~\cite{PhysRevResearch.3.013227}.
Although the approximated CD protocol is expected to be suitable for many-body systems by construction,
little efforts have been devoted so far to this direction.

In this paper, we apply the
variational CD protocol
to strongly correlated fermionic systems.
We formulate the variational optimization procedure
for general many-body Hamiltonians
in terms of a set of linear equations, where
its coefficients are set by the squared Frobenius norms of the nested commutators and
its solution vector gives the optimal variational parameters.
To treat multi-fermion-operator-product terms
in the nested commutators, we devise an exact numerical algorithm by
reassembling
fermionic creation and annihilation operators
into a special normal-ordered form,
which therefore
evades statistic errors in evaluating the optimal variational parameters
and escapes the formidable iterative matrix-vector multiplications
in evaluating the nested commutators.
This algorithm allows us to study systems up to
$L=14$ sites with $l\leqslant 3$
for a two-component fermionic Hubbard model at half filling.
By setting the ground state of a one-dimensional (1D) Hubbard model as the target state,
we examine whether the variational CD protocol can effectively
increase the fidelity for the ground-state preparation and
at the same time speed up the driving process.

The rest of this paper is organized as follows.
We briefly introduce the CD driving protocol
in Sec.~\ref{sec:cd_driving}
and the approximated AGP
in Sec.~\ref{sec:agp}.
Then, we formulate the variational CD protocol
in Sec.~\ref{sec:var_cd},
followed by a few remarks
in Sec.~\ref{sec:remarks}.
We describe our models and numerical
methods in Sec.~\ref{sec:MM} and
show numerical results in Sec.~\ref{sec:results}.
Finally, we conclude this paper in Sec.~\ref{sec:conclusion}.
We also provide some details of the formulation in Appendix~\ref{sec:proof}.
Throughout the paper, we set $\hbar = 1$.

\section{Formalism}\label{sec:form}

\subsection{CD driving}\label{sec:cd_driving}

Let us consider a time-dependent Hamiltonian $\hat{H}(t)$,
which has an instantaneous eigenstate
$|n(t)\rangle$ with energy $\epsilon_n(t)$,
i.e.,
\begin{align}
\label{eq:eq2}
\hat{H}(t)|n(t)\rangle = \epsilon_n(t)|n(t)\rangle.
\end{align}
According to the adiabatic theorem,
if the initial state at $t=0$ is an eigenstate,
it remains the corresponding instantaneous eigenstate
throughout the time evolution of the state,
provided that the evolution process is sufficiently slow
and the state remains non-degenerate~\cite{JPhysSocJpn.5.435}.
Then, the time-evolved state at later time $t$,
denoted by $|\psi_n(t)\rangle$,
can only differ from the instantaneous eigenstate $|n(t)\rangle$
by a phase factor~\cite{RevModPhys.82.1959}
\begin{align}
\label{eq:eq3}
|\psi_n(t)\rangle = e^{\text{i}
(\gamma_n^\text{geo} + \gamma_n^{\text{dyn}})  } |n(t)\rangle,
\end{align}
where
\begin{align}
\gamma_n^{\text{geo}}(t)&  = \text{i} \int_0^t \text{d}t^\prime  \langle n(t^\prime) |
\partial_{t^\prime} n (t^\prime) \rangle
\end{align}
and
\begin{align}
\gamma_n^{\text{dyn}}(t) & = - \int_0^t \text{d}t^\prime  \epsilon_n(t^\prime)
\end{align}
are the geometric and dynamic phases, respectively.

Notice that Eq.~(\ref{eq:eq3}) is only valid in the adiabatic regime.
We now intend to find a CD Hamiltonian $\hat{H}_{\text{CD}}(t)$
of which $|\psi_n(t)\rangle$ is the exact time-evolving state~\cite{JPhysAMathTheor.42.365303},
i.e.,
\begin{align}
\label{eq:eq5}
\text{i}\partial_t |\psi_n(t)\rangle = \hat{H}_{\text{CD}}(t) |\psi_n(t)\rangle.
\end{align}
In other words, $\hat{H}_{\text{CD}}(t)$ evolves in time
along the trajectories of the instantaneous eigenstates $|n(t)\rangle$
of $\hat{H}(t)$
[note that $|\psi_n(t)\rangle$ and $|n(t)\rangle$ are the same physical state].
Therefore, diabatic transitions among different eigenstates $|n'(t)\rangle$ for $n'\ne n$
are suppressed in this CD driving without the constraint of slow enough dynamics imposed by the adiabatic theorem.

From Eq.~(\ref{eq:eq3}),
the time-evolution operator $\hat{\cal U}(t)$
can be written as
\begin{align}
\label{eq:eq6}
\hat{\cal U}(t) = \sum_n e^{\text{i}( \gamma_n^{\text{geo}} + \gamma_n^{\text{dyn}})}
|n(t) \rangle \langle n(0) |,
\end{align}
which is apparently unitary.
To generate the dynamics of $\hat{\cal U}(t)$,
$\hat{H}_{\text{CD}}(t)$ needs to satisfy
\begin{align}
\label{eq:eq7}
\text{i} \partial_t \hat{\cal U}(t) = \hat{H}_{\text{CD}}(t) \hat{\cal U}(t).
\end{align}
Therefore,
\begin{align}
\label{eq:eq8}
\hat{H}_{\text{CD}}(t) = \text{i} (\partial_t \hat{\cal U}(t) ) \hat{\cal U}^\dag(t).
\end{align}
By inserting Eq.~(\ref{eq:eq6}) into Eq.~({\ref{eq:eq8}}), we obtain
\begin{align}
\label{eq:eq9}
\hat{H}_{\text{CD}}(t) = \hat{H}(t) + \hat{\mathcal{C}}(t)
\end{align}
with
\begin{align}
\hat{\mathcal{C}}(t) = \text{i} \sum_n \left( | \partial_t n \rangle  \langle n | - | n \rangle
\langle  n | \partial_t n  \rangle \langle n | \right).
\label{eq:eq10}
\end{align}
By substituting
\begin{equation}
\label{eq:eq11}
| \partial_t n \rangle = - \sum_{ m\,(\neq n) }
\frac{\langle m | \partial_t \hat{H} | n\rangle}{\epsilon_m - \epsilon_n}
|m\rangle
\end{equation}
into Eq.~(\ref{eq:eq10}),
we obtain the counter term $\hat{C}(t)$ given in Eq.~(\ref{eq:eq1}).

When the original Hamiltonian $\hat{H}(t)$
depends on time implicitly through
a driving function $\lambda(t)$,
i.e., $\hat{H}(t) = \hat{H}[\lambda(t)]$,
then the CD Hamiltonian is given as
\begin{align}
\label{eq:eq12}
\hat{H}_{\text{CD}} (t)
= \hat{H}(t)
+ \dot{\lambda}(t) \hat{\mathcal{A}}_{\lambda}(t),
\end{align}
where
\begin{align}
\label{eq:Ak}
\hat{\mathcal{A}}_{\lambda} (t)
= -\text{i} \sum_{m,n\,(m\ne n)} \frac{ |m\rangle\langle m |  \partial_{\lambda} \hat{H} | n\rangle \langle n| }{ \epsilon_m - \epsilon_n}
\end{align}
and $\dot{\lambda}(t)$ is the time derivative of $\lambda(t)$.
$\hat{\mathcal{A}}_{\lambda}(t)$ is called the adiabatic gauge potential (AGP),
which shares a similar structure
to the counter term $\hat{\mathcal{C}}(t)$
and encounters the same difficulties in dealing with many-body systems.
Hereafter, we refer to the original Hamiltonian
$\hat{H}(t)$ as the unassisted (UA) model and the CD Hamiltonian $\hat{H}_{\text{CD}} (t)$ as the CD model.

\subsection{Approximated AGP}\label{sec:agp}

To circumvent the aforementioned
difficulties,
an ansatz for approximating the AGP
has been proposed in Ref.~\cite{PhysRevLett.123.090602}.
The approximated AGP takes the following form:
\begin{alignat}{1}
\label{eq:A_kl}
\hat{\cal A}_\lambda^{(l)}(t)
&= \text{i} \sum_{k=1}^l \alpha_k(t) \hat{O}_{2k-1}(t),
\end{alignat}
where $l$ is the expansion order,
$\alpha_k(t)$ are real parameters to be determined,
and $\hat{O}_k(t)$ are nested commutators of the form
\begin{equation}
\label{eq:Ok}
\hat{O}_{k}(t)\equiv
[\underset{k}{\underbrace{\hat{H}(t),[\hat{H}(t),\cdots,[\hat{H}(t)}},\partial_\lambda \hat{H}(t)]]].
\end{equation}
Note that the operators $\hat{O}_{k}(t)$ can be defined recursively as
\begin{equation}
\label{eq:Ok2}
\hat{O}_{k}(t)=\left[\hat{H}(t),\hat{O}_{k-1}(t)\right],
\end{equation}
starting with $\hat{O}_0(t) \equiv \partial_\lambda \hat{H}(t)$.
By induction, it is easy to show that
$\hat{O}_{k}(t)$ is Hermitian (antihermtian) when $k$ is even (odd), i.e.,
\begin{alignat}{1}
\left[ \hat{O}_{k}(t) \right]^\dag &=(-1)^k\hat{O}_{k}(t). \label{eq:Ok_dag}
\end{alignat}
The antihermticity of $\hat{O}_{2k-1}(t)$ in Eq.~(\ref{eq:Ok_dag})
corroborates that $\hat{\cal A}_{\lambda}^{(l)}(t)$ is Hermitian.
The approximated AGP $\hat{\cal A}^{(l)}_\lambda(t)$ in Eq.~(\ref{eq:A_kl})
retrieves the exact AGP in Eq.~(\ref{eq:Ak}) in the limit $l\to \infty$.
In the following, for simplicity of notation, we omit to express the time dependence of quantities,
unless it is important to remind the dependence.

\subsection{Variational approach} \label{sec:var_cd}

It is suggested in Ref.~\cite{ProcNatlAcadSci.114.E3909} that
the optimal parameters $\alpha_k$ in Eq.~(\ref{eq:A_kl})
can be obtained by minimizing the action
\begin{equation}
\label{eq:Sl}
S_l \equiv
\left\langle \hat{G}_l, \hat{G}_l\right\rangle_{\text{F}}
= \left|\left|\hat{G}_l\right|\right|^2_{\text{F}},
\end{equation}
where
\begin{equation}
\label{eq:Gl}
\hat{G}_l
=\partial_\lambda \hat{H} - \text{i} \left[\hat{H},\hat{\cal A}^{(l)}_\lambda\right]
\end{equation}
is a Hermitian operator,
\begin{equation}
\label{eq:AB}
\left\langle \hat{A}, \hat{B} \right \rangle_{\text{F}} \equiv
\text{Tr}\left(\hat{A}^\dag \hat{B}\right)
\end{equation}
denotes the Frobenius (or Hilbert-Schmidt)
inner product of two operators $\hat{A}$ and $\hat{B}$,
and $||\hat{A}||_{\text{F}}=\sqrt{\langle \hat{A},\hat{A}\rangle_{\text{F}}}$
is the Frobenius norm.
By substituting Eq.~(\ref{eq:A_kl}) into Eq.~(\ref{eq:Gl})
and using Eq.~(\ref{eq:Ok2}), $\hat{G}_l$
can be expressed as a linear combination of
the even-order nested commutators
\begin{alignat}{1}
\hat{G}_l
=\hat{O}_0 + \sum_{k=1}^l \alpha_k \hat{O}_{2k}
=\sum_{k=0}^l \alpha_k \hat{O}_{2k},
\end{alignat}
where $\alpha_0\equiv 1$.
The minimization condition with respect to $\alpha_k$ thus reads
\begin{equation}
\label{eq:Sla}
\frac{\partial S_l}{\partial \alpha_k}
= 2 \sum_{m=0}^l \alpha_m
\left\langle \hat{O}_{2m}, \hat{O}_{2k}\right\rangle_{\text{F}} = 0
\end{equation}
for $1 \leqslant k \leqslant l$.

Note that, as shown in Appendix~\ref{sec:proof}, the inner product in Eq.~(\ref{eq:Sla}) can be simplified as
\begin{alignat}{1}
\label{eq:OO}
\left\langle \hat{O}_{2m}, \hat{O}_{2k}\right\rangle_{\text{F}}
= \left|\left|\hat{O}_{m+k}\right|\right|_{\text{F}}^2.
\end{alignat}
To further simplify the notation, we denote
the squared Frobenius norm of the nested commutator as
\begin{equation}
\label{eq:gamma_k}
\Gamma_{k}\equiv \left|\left|\hat{O}_k \right|\right|_{\text{F}}^2.
\end{equation}
Then, the minimization condition in Eq.~(\ref{eq:Sla}) is now given as
\begin{equation}
\label{eq:linear}
\sum_{m=0}^l \alpha_m \Gamma_{m+k} = 0.
\end{equation}
Recalling that $\alpha_0=1$,
Eq.~(\ref{eq:linear}) can be written finally as a set of linear equations
\begin{equation}
\begin{bmatrix}
\Gamma_2 & \Gamma_3 & \cdots & \Gamma_{l+1} \\
\Gamma_3 & \Gamma_4 & \cdots & \Gamma_{l+2} \\
\vdots & \vdots & \ddots & \vdots \\
\Gamma_{l+1} & \Gamma_{l+2} & \cdots & \Gamma_{2l}
\end{bmatrix}
\begin{bmatrix}
\alpha_1 \\
\alpha_2 \\
\vdots \\
\alpha_l
\end{bmatrix}
=
-
\begin{bmatrix}
\label{eq:linear2}
\Gamma_1 \\
\Gamma_2 \\
\vdots \\
\Gamma_l
\end{bmatrix}.
\end{equation}
Therefore, regardless of the number $l$ of variational parameters $\{\alpha_k\}_{k=1}^l$,
the optimal parameters are simply obtained
deterministically as the solution vector of Eq.~(\ref{eq:linear2}).
In particular, in the $l = 1$ case, the optimal parameter is
$\alpha_1 = -\Gamma_1 / \Gamma_2 < 0$.

\subsection{Remarks on Eq.~(\ref{eq:linear2}) }
\label{sec:remarks}

Here, we give three remarks on Eq.~(\ref{eq:linear2}).
First,
the $l\times l$ symmetric matrix in the left-hand side of Eq.~(\ref{eq:linear2})
is a Gram matrix whose $(i,j)$ entry is given by
the inner product
\begin{equation}
\label{eq:eq27}
\Gamma_{i+j} =
\left|\left|\hat{O}_{i+j} \right|\right|_{\text{F}}^2
= \left \langle \hat{O}_{2i},\hat{O}_{2j} \right \rangle_{\text{F}},
\end{equation}
and hence the matrix is positive semidefinite.
To be more specific,
let us define a $D^2 \times l$ matrix
$\boldsymbol{O}\equiv [\boldsymbol{o}_{2}, \ \boldsymbol{o}_{4}, \ \cdots, \ \boldsymbol{o}_{2l}]$
with
$D$ being the dimension of the Hilbert space and
$\boldsymbol{o}_{2k}$ being $D^2$ dimensional vector
whose elements are given by an arbitrary sequence of
$\{\{\langle e_m |\hat{O}_{2k} | e_n\rangle\}_{m=1}^D\}_{n=1}^D$
for any set of orthonormalized basis $\{|e_m\rangle\}_{m=1}^D$
such that
$\boldsymbol{o}_{2i}^\dag \boldsymbol{o}_{2j}
=\text{Tr}\left(\hat{O}_{2i}^\dag \hat{O}_{2j}\right)=\Gamma_{i+j}$.
Hence, the $l\times l$ matrix in the left-hand side of Eq.~(\ref{eq:linear2}), now denoted as $\boldsymbol{\Gamma}$,
can be written as $\boldsymbol{\Gamma}=\boldsymbol{O}^\dag \boldsymbol{O}$.

Second,
by using the eigenpairs $\{\epsilon_n, |n\rangle\}$ of $\hat{H}$
and the identity
$\langle m | \hat{O}_k |n \rangle=
\omega_{mn}
\langle m|\hat{O}_{k-1}|n\rangle$
with $\omega_{mn}\equiv(\epsilon_m-\epsilon_n)$,
we can show that
\begin{equation}
\label{eq:eq28}
\Gamma_k
=\sum_{m,n}\left|\left\langle m\left| \hat{O}_k\right| n \right\rangle \right|^2
=\sum_{m,n}\left|\left\langle m\left| \partial_\lambda \hat{H}\right| n \right\rangle \right|^2
\omega_{mn}^{2k}.
\end{equation}
Equation~(\ref{eq:eq28}) implies that $\Gamma_k$
coincides with the quantity $\Gamma_\lambda^{(k)}$
defined in the supplemental material of Ref.~\cite{PhysRevLett.123.090602}, where
$\Gamma_\lambda^{(k)}$ is called the $2k$th moment of a response function
$ \Gamma_\lambda(\omega)\equiv
\sum_{mn}\left|\langle m\left| \partial_\lambda \hat{H}\right| n \rangle \right|^2
\delta(\omega-\omega_{mn})$
, i.e.,
$\Gamma_\lambda^{(k)}
=\int \text{d} \omega \Gamma_\lambda(\omega)\omega^{2k}$.

Third,
as anticipated from the fact that $\hat{O}_k$ contains
the Hamiltonian powers of order $k$,
or also from Eq.~(\ref{eq:eq28}),
$\Gamma_k$ may increase exponentially in $k$.
Consequently,
in order to satisfy Eq.~(\ref{eq:linear2}),
$|\alpha_k|$ is expected to
decrease exponentially in $k$.
We will discuss this point later in detail
with our numerical results in Sec.~\ref{sec:results}.

\section{Models and Methods}
\label{sec:MM}

\subsection{ Models } \label{sec:model}

\subsubsection{UA model}

As the UA model, we consider the following time-dependent two-component fermionic Hubbard model~\cite{Hubbard1963}
on a 1D chain consisting of $L$ sites under open-boundary conditions:
\begin{equation}
\label{eq:hubbard}
\hat{H}(t) = \hat{H}[\lambda(t)]= \hat{H}_J + \lambda(t) \hat{H}_{U},
\end{equation}
where
\begin{equation}
\label{eq:kim}
\hat{H}_{J}=
-J\sum_{\langle i,j \rangle}\sum_{\sigma=\uparrow,\downarrow}
\left(\hat{c}_{i\sigma}^\dag \hat{c}_{j\sigma} + {\text{ H.c.}}\right)
\end{equation}
is the hopping part and
\begin{equation}
\label{eq:Hu}
\hat{H}_{U}=
U\sum_{i}
\hat{n}_{i\uparrow} \hat{n}_{i\downarrow}
\end{equation}
is the interacting part.
Here,
$\hat{c}_{i\sigma}$
($\hat{c}_{i\sigma}^\dag$)
is the annihilation (creation) operator
of a fermion at site $i$ with spin $\sigma\,(=\uparrow,\downarrow)$ and
$\hat{n}_{i\sigma}=\hat{c}_{i\sigma}^\dag \hat{c}_{i\sigma}$
is the fermion density operator.
$J$ is the hopping amplitude,
$U\geqslant 0$ is the strength of the on-site interaction,
and the sum $\sum_{\langle i,j \rangle}$ in Eq.~(\ref{eq:kim})
runs over all pairs of nearest-neighbor sites $i$ and $j$ on a 1D lattice under open-boundary conditions.

Here, we adopt the same driving function $\lambda(t)$ as in Ref.~\cite{PhysRevLett.123.090602}, i.e.,
\begin{equation}
\label{eq:lambda}
\lambda(t)=\sin^2\left[  \frac{\pi}{2} \sin^2\left(\frac{\pi t}{2T}\right)\right],
\end{equation}
satisfying $\lambda(0) = 0$ at the initial time $t=0$
and $\lambda(T) = 1$ at the finial time $t=T$ of the driving
(see Fig.~\ref{fig:lambda} in Ref.~\cite{SM}).
Therefore, $\hat{H}(t=0)=\hat{H}_J$ is simply the noninteracting tight-binding model and
$\hat{H}(t=T)=\hat{H}_J+\hat{H}_U$ is the desired Hubbard model $\hat{H}_{\rm HB}$.
Note that the parameter $T$ in Eq.~(\ref{eq:lambda}) represents the driving period and
characterizes the driving rate of the dynamic process;
the smaller $T$ corresponds to the faster driving, while
the limit $T \to \infty$ corresponds to the perfect adiabatic driving.
We should also note that obviously $\hat{H}(t)$ in
Eq.~(\ref{eq:hubbard}) preserves the global U(1) symmetries for
the spin and the charge sectors as in the standard time-independent Hubbard
model $\hat{H}_{\rm HB}$~\cite{Hubbard1963,essler2005}, and thus the total number $N_{f}$ of fermions and the $z$ component
$S_z$ of the total spin of fermions are both good quantum numbers.
In this paper, we set a unit of energy (time) to be $J$ ($1/J$),

\subsubsection{CD model}

The CD Hamiltonian $\hat{H}_{\text{CD}}^{(l)}(t)$ with the $l$th order
approximated AGP is obtained by replacing
$\hat{\cal A}_\lambda(t)$ in Eq.~(\ref{eq:eq12})
with $\hat{\cal A}_{\lambda}^{(l)}(t)$, i.e.,
\begin{align}
\label{eq:hcd}
\hat{H}_{\text{CD}}^{(l)}(t) = \hat{H}(t) + \dot{\lambda}(t) \hat{\mathcal{A}}^{(l)}_{\lambda}(t),
\end{align}
and now it has acquired the superscript $l$ to represent explicitly the order of the approximation.
The time derivative of the driving function is given by
\begin{equation}
\label{eq:lambda_t}
\dot{\lambda}(t)=\frac{\pi^2}{4T} \sin\left(\frac{\pi t}{T}\right)
\sin\left[\pi\sin^2\left(\frac{\pi t}{2T}\right)\right],
\end{equation}
which satisfies
$\dot{\lambda}(0) = \dot{\lambda}(T) = 0$ at the initial and final times
(see Fig.~\ref{fig:lambda} in Ref.~\cite{SM})
and hence ensures that the CD and UA models
coincide at the beginning
and the end of the evolution at $t=0$ and $t=T$, respectively.

We note that $\hat{O}_0(t)$
and $\hat{O}_1(t)$
are given by
$\hat{O}_0(t) = \partial_\lambda \hat{H}[\lambda(t)] = \hat{H}_U$,
$\hat{O}_1(t) = [\hat{H}_J, \hat{H}_U]$, both thus being time independent, and
the time dependence of $\hat{O}_k(t)$ appears only for $k\geqslant2$.
Moreover, it is important to notice that the CD Hamiltonian $\hat{H}_{\rm CD}^{(l)}(t)$ in Eq.~(\ref{eq:hcd})
also preserves the global U(1) symmetries for
the spin and the charge sectors, and therefore the total number $N_{f}$ of fermions and the $z$ component
$S_z$ of the total spin of fermions are also both good quantum numbers.
Since the $\hat{O}_k(t)$ operators also preserve these symmetries, the trace operation necessary to
evaluate the squared Frobenius norm for $\Gamma_k$ in Eq.~(\ref{eq:gamma_k}) is limited within the subspace
of the whole Hilbert space.

\subsection{Methods }

\subsubsection{ Time evolution }

Numerically, the time-evolved state is calculated through
\begin{equation}
\label{eq:psi_t}
|\psi (t+\Delta t) \rangle = \text{e}^{-\text{i} \hat{H}(t) \Delta t} |\psi (t)\rangle,
\end{equation}
where $\Delta t = T/N_T$ is a small time interval
and $N_T$ is the number of time steps.
Although here we assume that the dynamics is governed by the UA Hamiltonian $\hat{H}(t)$, the following argument is
similarly applied to the dynamics driven by the approximated CD Hamiltonian $\hat{H}_{\rm CD}^{(l)}(t)$.
For small systems,
we treat the time-evolution operator $ \text{e}^{-\text{i} \hat{H}(t) \Delta t}$ exactly by the full diagonalization method.
For large systems, we expand the time-evolution operator
by the Chebyshev polynomials as~\cite{JChemPhys.81.3967,Weisse2008}
\begin{align}
\label{eq:expand}
\text{e}^{-\text{i} \hat{H}(t) \Delta t}
&= \text{e}^{-\text{i}(a \hat{\tilde{H}}(t)+b)\Delta t} \notag \\
&=  \text{e}^{-\text{i}b\Delta t} \left(c_0 + 2 \sum_{k=1}^N c_k T_k(\hat{\tilde{H}}(t)) \right)
+ \mathcal{O}( (a\Delta t)^{N+1} ),
\end{align}
where
$\hat{\tilde{H}}=(\hat{H}-b)/a$ is
the scaled Hamiltonian and the parameters
$a$ and $b$ are chosen so as to satisfy
$\rho( \hat{\tilde{H}}) \leqslant 1$
with $\rho(\cdot)$ denoting the spectral radius.
$T_k$ is the $k$th-order Chebyshev polynomial of the first kind,
$N$ is the expansion order,
and the expansion coefficients $c_k$ for $k\geqslant0$ are given by~\cite{Weisse2008}
\begin{align}
\label{eq:ck}
c_k = \int_{-1}^{1} \frac{T_k(x) \text{e}^{-\text{i}ax\Delta t}}{ \pi \sqrt{1-x^2} } \text{d}x = (-\text{i})^k J_k(a\Delta t).
\end{align}
Here $J_k$ denotes the $k$th-order Bessel function of the first kind.

We have confirmed that
$\Delta t=0.001$ and $N=10$ are enough to reproduce
the time evolution of fidelity [defined in Eq.~(\ref{eq:fidelity})] obtained
by the full diagonalization method for small $L\leqslant 8$
(with Hilbert space dimension $D\leqslant 4900$)
within $10^{-6}$ (see Fig.~\ref{fig:check} in Ref.~\cite{SM}).
For larger systems, we have checked that the results are already well converged with $N=10$ for $\Delta t=0.001$
by comparing the results with different values of $N$. Therefore, we set $\Delta t=0.001$ and $N=10$
to obtain the numerical results shown in
Sec.~\ref{sec:results}.

\subsubsection{ Construction of CD model } \label{sec:const_cd}

In order to evolve the state $|\psi(t)\rangle$ in time via
the CD Hamiltonian $\hat{H}_{\text{CD}}^{(l)}(t)$,
we need to calculate $\alpha_k$ at each time step from
the squared Frobenius norm $\Gamma_k$ ($k=1,\ldots,2l$)
of the nested commutators $\hat{O}_k$ by solving the set of linear equations in Eq.~(\ref{eq:linear2}).
In addition, we have to operate $\{\hat{O}_{2k-1}\}_{k=1}^l$ to $|\psi(t)\rangle$
at each time step.
The most straightforward way to do the latter
for large systems ($D>4900$) is to
evaluate $\hat{O}_{2k-1} |\psi(t)\rangle$ from
$\hat{H}\hat{O}_{2k-2}|\psi(t)\rangle$ and
$\hat{O}_{2k-2}\hat{H}|\psi(t)\rangle$,
based on the recursive formula in Eq.~(\ref{eq:Ok2}).
By applying the same procedure of random-phase vectors as in Ref.~\cite{Iitaka2004},
$\Gamma_k={\rm Tr}(\hat{O}_k^\dag \hat{O}_k)$
can also be estimated
without explicitly constructing matrix representations for $\hat{O}_k$,
similar to the evaluation of
Frobenius inner products of many-body operators in Ref.~\cite{seki2021}.
Accordingly, statistical errors in $\Gamma_k$ due to the samplings are
inevitable in this approach, although it is expected to become smaller for larger $D$.

Instead,
we devise an exact algorithm from a constructive approach
to treat large systems, avoiding
the recursive operation of $\hat{O}_k$ operators and
the statistical samplings for $\Gamma_k$.
The main strategy is simply to implement the analytic representations
of the $\hat{O}_k$ operators for $k=0,1,\ldots,2l$,
which consist of many multi-fermion-operator-product terms of the form
\begin{align}
\hat{h}_{m,x}(t) =
\lambda(t)^{m} \beta_x \hat{c}^\dag_i \hat{c}^\dag_j \cdots \hat{c}_p \hat{c}_q \cdots.
\end{align}
Here,
$m$ is a time-independent integer number,
$\beta_x$ is a time-independent complex coefficient,
$x\equiv\{i,j,\cdots,p,q,\cdots\}$
denotes a set of indexes,
and $i,j,\cdots,p,q,\cdots$ are collective indexes that label
the site and spin of fermions.
Note that the time dependence occurs only through $\lambda(t)^{m}$ and
particularly, in our case,
$\hat{O}_0$ and $\hat{O}_1$ are time independent with $m = 0$.
Since the hopping part
$\hat{H}_J$ has $m = 0$
and the interacting part $\lambda \hat{H}_U$
has $m=1$, it is easy to verify
that the maximal $m$ for $\hat{O}_k$ with $1\leqslant k \leqslant 2l$ is $k-1$.
Namely, $\hat{O}_k$ can be expressed as
\begin{align}
\hat{O}_k(t)
= \sum_{m, x} \hat{h}_{m, x}(t)
= \sum_{m = 0}^{k-1} \lambda(t)^m \sum_{x} \beta_x
c^\dag_i c^\dag_j \cdots c_p c_q \cdots.
\end{align}
Therefore, $\Gamma_k$ is given by
\begin{align}
\Gamma_k & =
\text{Tr}\left( \hat{O}_k^\dag \hat{O}_k \right)   \nonumber \\
& = \sum_{m = 0}^{k-1} \lambda(t)^{2m}  \text{Tr} \left( \sum_x \beta_x^{(k,m)}
c^\dag_i c^\dag_j \cdots c_p c_q \cdots \right) \nonumber \\
& \equiv
\sum_{m = 0}^{k-1} \lambda(t)^{2m} S^{k,m},
\label{eq:Gamma_k2}
\end{align}
where we indicate the $k$ and $m$ dependencies of $\beta_x$ explicitly by $\beta_x^{(k,m)}$
and $S^{k,m}\equiv  \text{Tr} \left( \sum_x \beta_x^{(k,m)}
c^\dag_i c^\dag_j \cdots c_p c_q \cdots \right)$
constitutes a $2l \times 2l$ lower triangular matrix
with $ 1 \leqslant k \leqslant 2l$ and $ 0 \leqslant m \leqslant 2l-1$
(see Ref.~\cite{SM} for a concrete example).
Notice that the terms with odd power index $2m+1$ of $\lambda(t)$ are
absent in Eq.~(\ref{eq:Gamma_k2}) because
they do not contribute to the trace in our model on a bipartite lattice.
Since $S^{k,m}$ is time independent, we can evaluate $S^{k,m}$ once prior to the time evolution
and use them repeatedly to evaluate $\Gamma_k$ at each time step.
By doing so, the computational cost can be greatly reduced (see Ref.~\cite{SM} for details).

To further facilitate the numerical manipulation, we implement operations among multi-fermion-operator-product
terms $\{\hat{h}_{m,x}\}$ for addition of terms,
scalar multiplication, and
commutation of two terms
that takes into account the canonical anticommutation relations
$\{\hat{c}_i, \hat{c}_j^\dag\} = \delta_{ij}$ and
$\{\hat{c}_i, \hat{c}_j\} = 0$, and then reassemble the terms $\{\hat{h}_{m,x}\}$ into
a special normal-ordered form, where the collective indexes
$x = \{i,j,\cdots, p,q,\cdots\}$ are ordered for faster numerical treatment.
Before starting the time evolution, we construct analytic representations for
the $\hat{O}_k$ operators and
the diagonal terms of $\hat{O}_k^\dag \hat{O}_k$, from which
the time-independent $S^{k,m}$
are evaluated by tracing these diagonal terms over
all the bases in the Hilbert pace.
At each time step, $\Gamma_k$ can now be easily calculated
from Eq.~(\ref{eq:Gamma_k2})
and the optimal variational parameters $\alpha_k$ is obtained
by solving the set of linear equations in Eq.~(\ref{eq:linear2}).
Once we determine the optimal $\alpha_k$,
an analytical representation of the CD model $\hat{H}_{\text{CD}}^{(l)}(t)$ in Eq.~(\ref{eq:hcd})
is obtained by adding the AGP to the UA model.
We should emphasis that $\Gamma_k$ for $1 \leqslant k \leqslant 2l$ are
evaluated without introducing any random samplings and hence it is
free of statistical errors.
More details on this constructive approach are described in Ref.~\cite{SM}.

\begin{figure}
\includegraphics[width=.9\columnwidth]{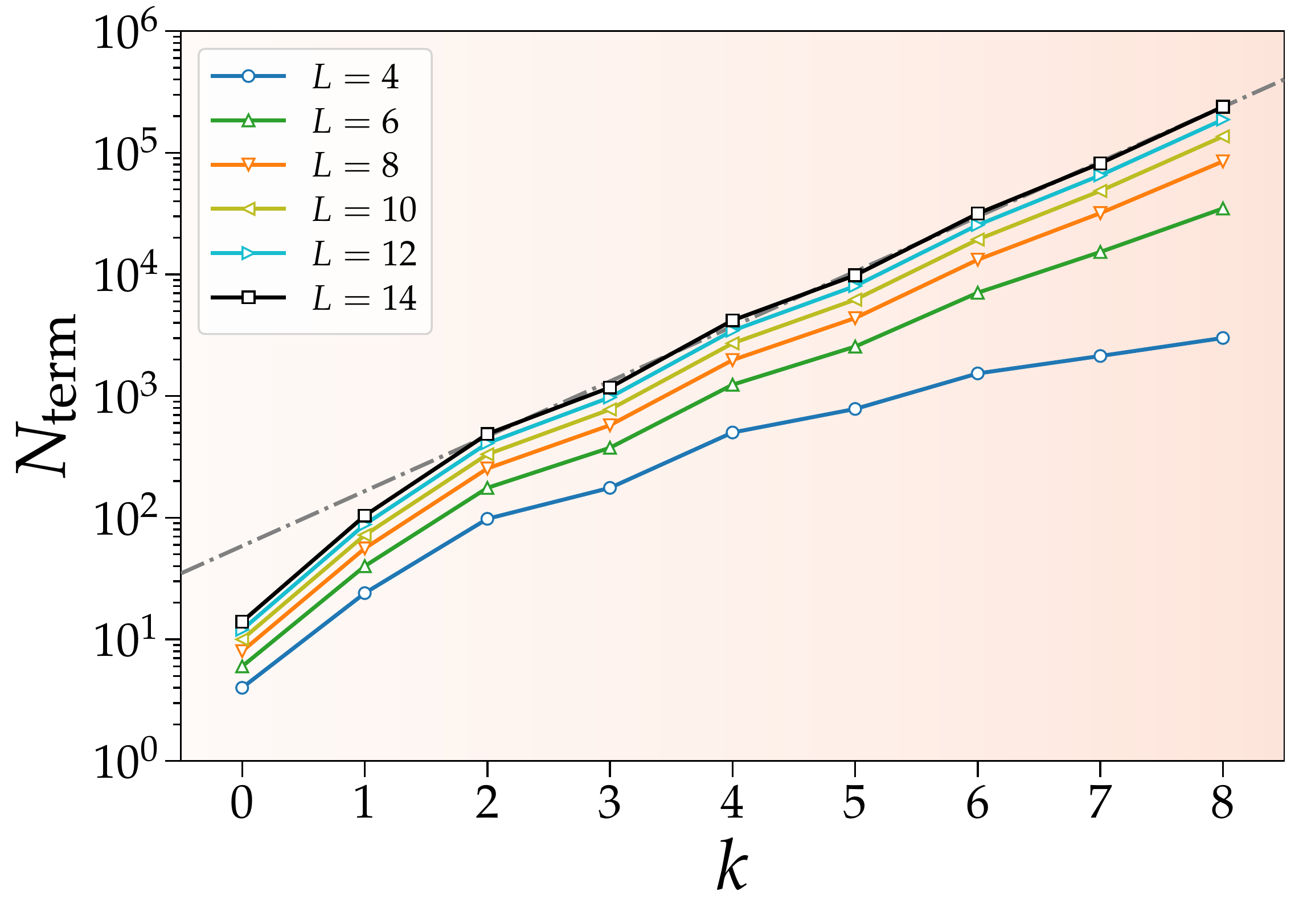}
\caption{
The number $N_{\rm term}$ of terms in the $\hat{O}_k$ operators
as a function of $k$ for various system sizes $L$.
The grey dash-dotted line
indicates a line of $N_{\rm term} = 10^{a k + b}$
with $a = 0.4512$ and $b = 1.7678$.
}
\label{fig:term}
\end{figure}

Finally, we remark a caveat to the constructive approach.
As shown in Fig.~\ref{fig:term},
the number $N_{\rm term}$ of terms in $\hat{O}_k$
increases exponentially in $k$.
This is simply a reflection of the fact that
the $\hat{O}_k$ operators induce long-range
multi-body interacting terms to the CD model.
As a consequence,
the number of terms in $\hat{H}_{\text{CD}}^{(l)}(t)$
also grows exponentially with the expansion order $l$.
Therefore, the feasibility of the constructive
approach is primarily determined by the expansion order $l$.
As shown in Sec.~\ref{sec:results}, we are able to study
systems up to $L = 14$ sites
~($D \leqslant 11778624 $)
with $l \leqslant 3$ under open-boundary conditions.

\section{Numerical Results}
\label{sec:results}

\subsection{ Physical quantities }

Let us first summarize quantities calculated in our numerical simulations.
Our primary concern is the time evolution of fidelities defined by
\begin{align}
\label{eq:fidelity}
\begin{split}
F_{tt} & = |\langle n(t) | \psi(t) \rangle |^2,    \\
F_{0t} & = |\langle n(0) | \psi(t) \rangle |^2,   \\
F_{Tt} & = |\langle n(T) | \psi(t) \rangle |^2.  \\
\end{split}
\end{align}
Here $F_{tt}$ is the overlap between
the instantaneous eigenstate $|n(t) \rangle$ of the UA model $\hat{H}(t)$,
and the time-evolved state $|\psi(t) \rangle$ driven by either the UA or CD model.
Accordingly, $F_{0t}$
($F_{Tt}$) is the overlap between the instantaneous eigenstate of the initial (final) Hamiltonian
and the time-evolved state.
In this study,
the initial state at $t=0$ is set to be the ground state of $\hat{H}(t=0)=\hat{H}_{\rm CD}^{(l)}(t=0)$,
and thus $|\psi(t=0)\rangle=|n(t=0)\rangle$, which is nothing but the ground state of the noninteracting fermions.
At the final time of the evolution, i.e., at $t=T$, the UA and CD models are again the same,
$\hat{H}(t=T)=\hat{H}_{\rm CD}^{(l)}(t=T)$, and $|n(t=T) \rangle$ is the ground state of the Hubbard model
$\hat{H}_{\rm HB}$.
The fidelity $F_{tt}$ at $t=T$, denoted simply as $F_{TT}$, characterizes
how faithfully we have prepared the ground state of the target Hubbard model.

In addition, we examine
the time evolution of the lowest two eigenvalues, $E_0$ and $E_1$, of the UA and CD models
with the associated spectrum gap $\Delta E = E_1 - E_0$,
the variational parameters $\{\alpha_k\}$, and
the magnitudes of AGP terms, $\{M_k\}$,
quantified by
$M_k = \dot{\lambda} |\alpha_k| \cdot || \hat{O}_{2k-1} ||_\text{F}$.

\subsection{System-size dependence at half filling }

In accordance with the global U(1) symmetries for the charge and the spin sectors
of the UA and CD models discussed in Sec.~\ref{sec:model},
we compute the time evolution in each sector
with the fixed numbers of up and down fermions, $N_\uparrow$ and $N_\downarrow$, respectively.
Therefore, for even number $N_{f}$ of fermions,
the sector
with $N_{\uparrow} = N_{\downarrow} = N_{f}/2$,
i.e., the total spin $S_z=0$, has the largest Hilbert space dimension
$D = \binom{L}{N_{\uparrow}} \binom{L}{N_{\downarrow}}$.
In this section, we focus on half filling, i.e., $N_{f}=L$, at which it is known that $S_z=0$ for the ground state of $\hat{H}(t=0)$
as well as $\hat{H}(t=T)$ for even $L$ under open-boundary conditions.

\begin{figure*}
\includegraphics[width=\hsize]{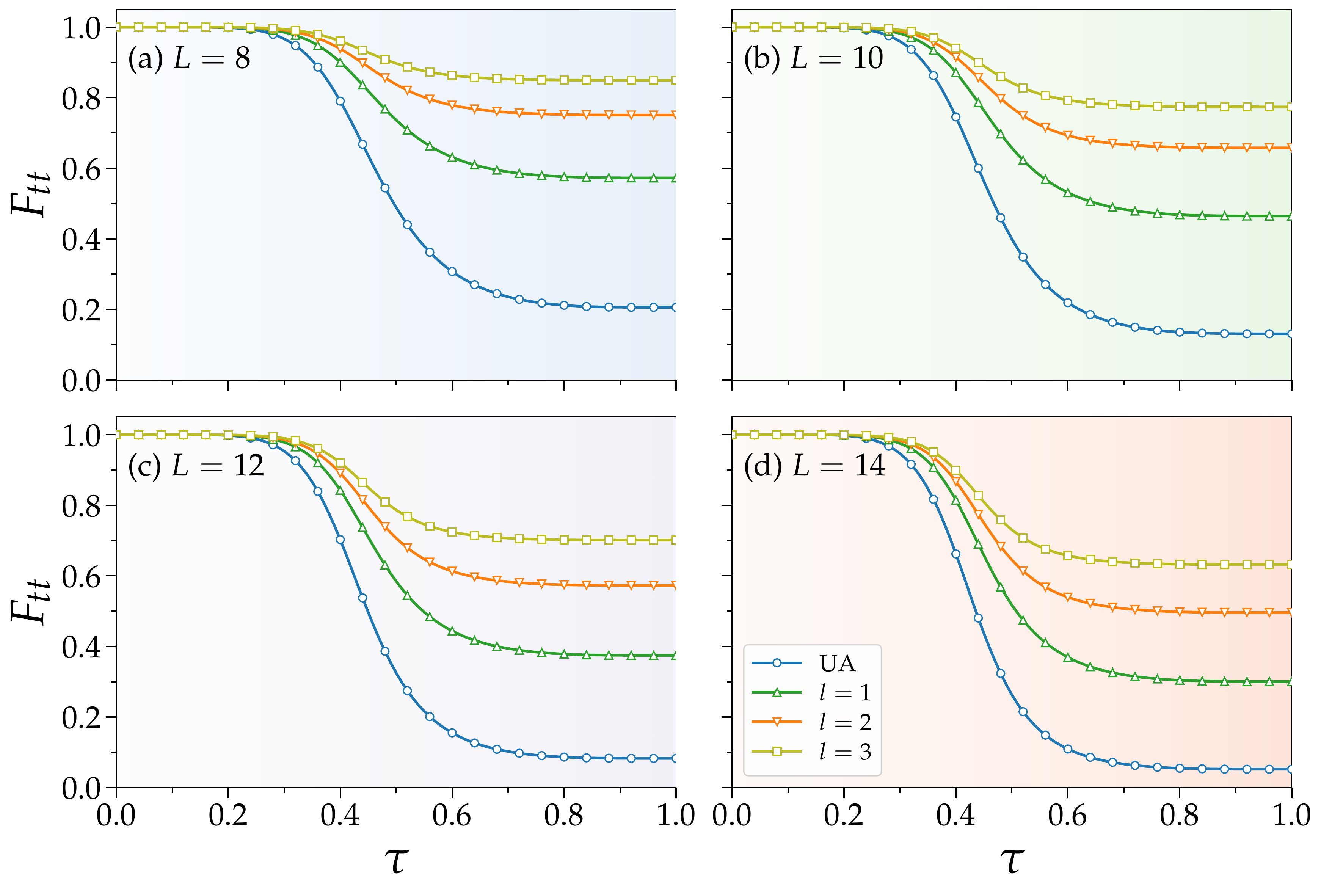}
\caption{
The time evolution of fidelity $F_{tt}$ for the UA model
and the CD models with different driving orders $l = 1,2,3$
on the 1D chains of
(a) $L = 8$,
(b) $L = 10$,
(c) $L = 12$ and
(d) $L = 14$ sites at half filling.
The remaining parameters are $U=8$, $T=0.1$, $N_T = 100$, $\Delta t = T/N_T=0.001$, and $N=10$.
Here, time $t$ in the horizontal axis is scaled by the driving period $T$, i.e., $\tau=t/T$.
}
\label{fig:fidelity}
\end{figure*}

Figure~\ref{fig:fidelity} shows the time evolution of fidelity $F_{tt}$
with respect to the scaled time $\tau = t/T$
for the UA and CD ($l=1,2,3$) models with $U=8$ and different system sizes
$L = 8, 10$, $12$, and $14$.
Here, we choose the driving period $T = 0.1$, $N_T = 100$ with $\Delta t= T/N_T = 0.001$,
and the order of the Chebyshev polynomial expansion $N = 10$. This set of
parameters guarantees converged results~\cite{SM}.
It is clearly observed in Fig.~\ref{fig:fidelity} that, for each $L$, increasing $l$ remarkably improves the
fidelity during the evolution.
In the UA model, the final fidelities $F_{TT}$ at $t=T$
are 20.59\%, 13.10\%, 8.27\%, and 5.19\% for $L = 8,10,12$, and $14$, respectively.
These fidelities rapidly increase to
57.27\%, 46.47\%, 37.46\%, and 30.04\% when the first-order CD protocol is employed.
The highest $F_{TT}$ achieved in the third-order CD protocol
are 84.93\%, 77.41\%, 70.11\%, and 63.20\% for $L = 8,10,12$, and $14$, respectively.
Although we cannot continue to increase the order $l$ of the CD protocol any longer for these system sizes
because of the computational cost,
even better fidelity is expected for the higher-order CD protocol, as demonstrated
in Fig.~\ref{fig:fidelity2} for smaller systems sizes~\cite{SM}.

\begin{figure*}
\includegraphics[width=\hsize]{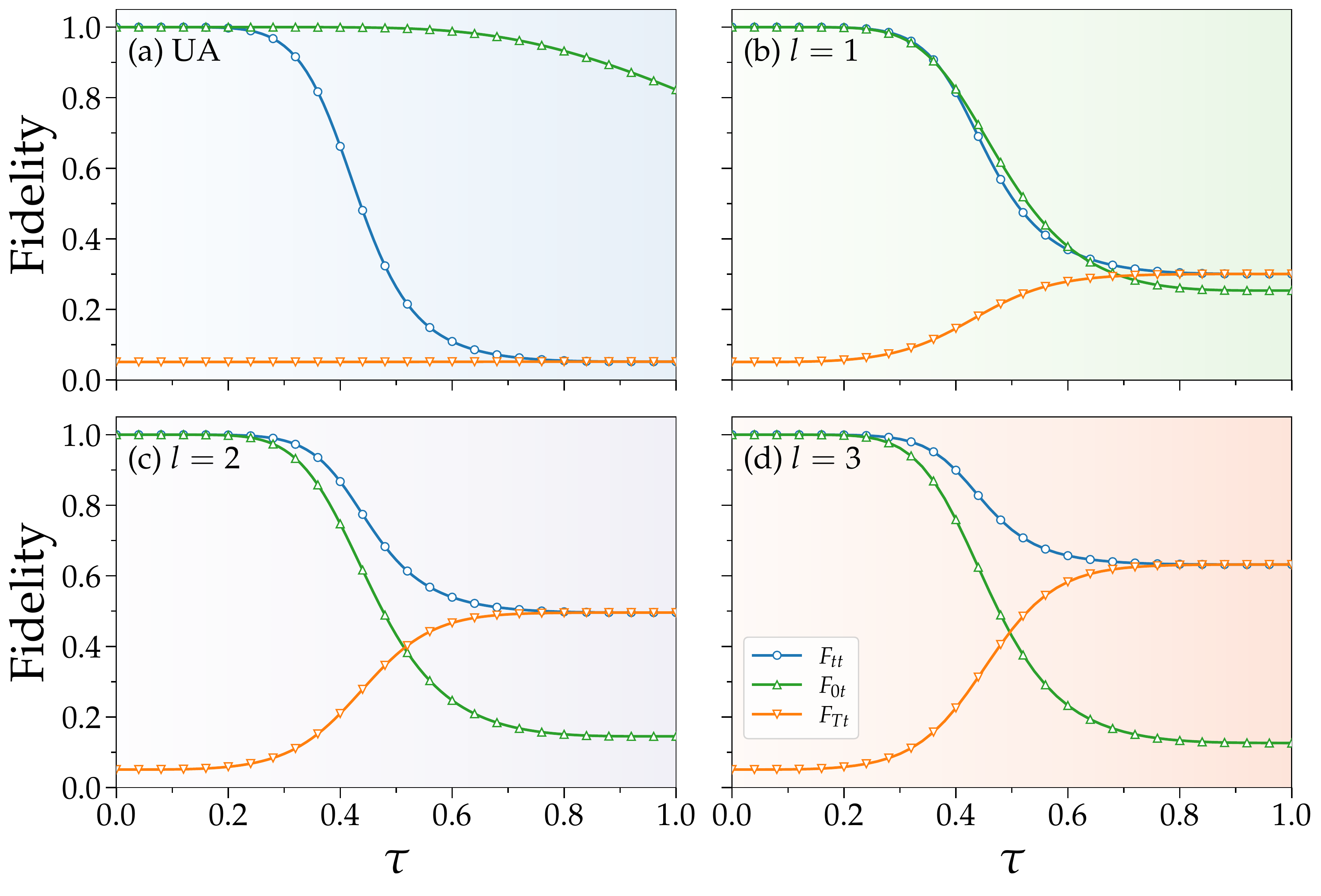}
\caption{
The time evolution of fidelities $F_{tt}$, $F_{0t}$, and $F_{Tt}$
for (a) the UA model and the CD models with (b) $l=1$, (c) $l=2$
and (d) $l = 3$
on the 1D chain of
$L = 14$ sites at half filling.
The remaining parameters are the same as in Fig.~\ref{fig:fidelity}.
Note that the results for $F_{tt}$ are exactly the same as those in Fig.~\ref{fig:fidelity}(d).
}
\label{fig:fidelity3}
\end{figure*}

Figure~\ref{fig:fidelity3} shows the time evolution of
the three different fidelities $F_{tt}, F_{0t}$, and $F_{Tt}$
for the UA and CD ($l = 1,2,3$) models on $L=14$ sites.
In the UA case [Fig.~\ref{fig:fidelity3}(a)],
$F_{0t}$ is very close to one for the first half of the driving period, i.e., $\tau < 0.5$,
and it remains to be a large value, as large as 82.27\%,
even at the end of driving period, i.e, $t = T$.
On the other hand,
$F_{Tt}$ is nearly constant throughout the whole driving period,
i.e., $F_{Tt} \approx F_{TT}\,(=5.19$\%)
[see the orange line in Fig.~\ref{fig:fidelity3}(a)].
These results suggest that
the time-evolved state $|\psi(t)\rangle$ for the UA model
remains very close to the initial state $|n(0)\rangle$
until the end of the driving and,
although $F_{tt}$ smoothly connects $F_{0t}$ at $t = 0$ to $F_{Tt}$ at $t = T$,
the time dependence of $F_{tt}$
arises essentially through the time dependence of $|n(t)\rangle$ itself, rather than $|\psi(t)\rangle$.
Therefore, the UA model with the driving period $T = 0.1$ is within
an impulse regime where the driving period is too short for the system to response~\cite{NewJPhys.12.093025}.
As shown in Figs.~\ref{fig:fidelity3}(b)-\ref{fig:fidelity3}(d), when the CD driving protocols are employed,
the fidelity $F_{Tt}$ increases with time and reaches to significantly large values at the end of the driving period.
Accordingly, the fidelity $F_{0t}$ dramatically decreases as the state $|\psi(t)\rangle$ evolves in time.
The CD driving protocol by definition defies
that
the system is in an impulse regime, as demonstrated here for the cases with the driving period $T=0.1$.
In fact, in the limit $l \to \infty$,
the perfect final fidelity is anticipated,
regardless of how short the driving period $T$ is
(as a demonstration, see Fig.~\ref{fig:fidelity4} in Ref.~\cite{SM}
for a smaller system with $L = 6$).

\begin{figure*}
\includegraphics[width=\hsize]{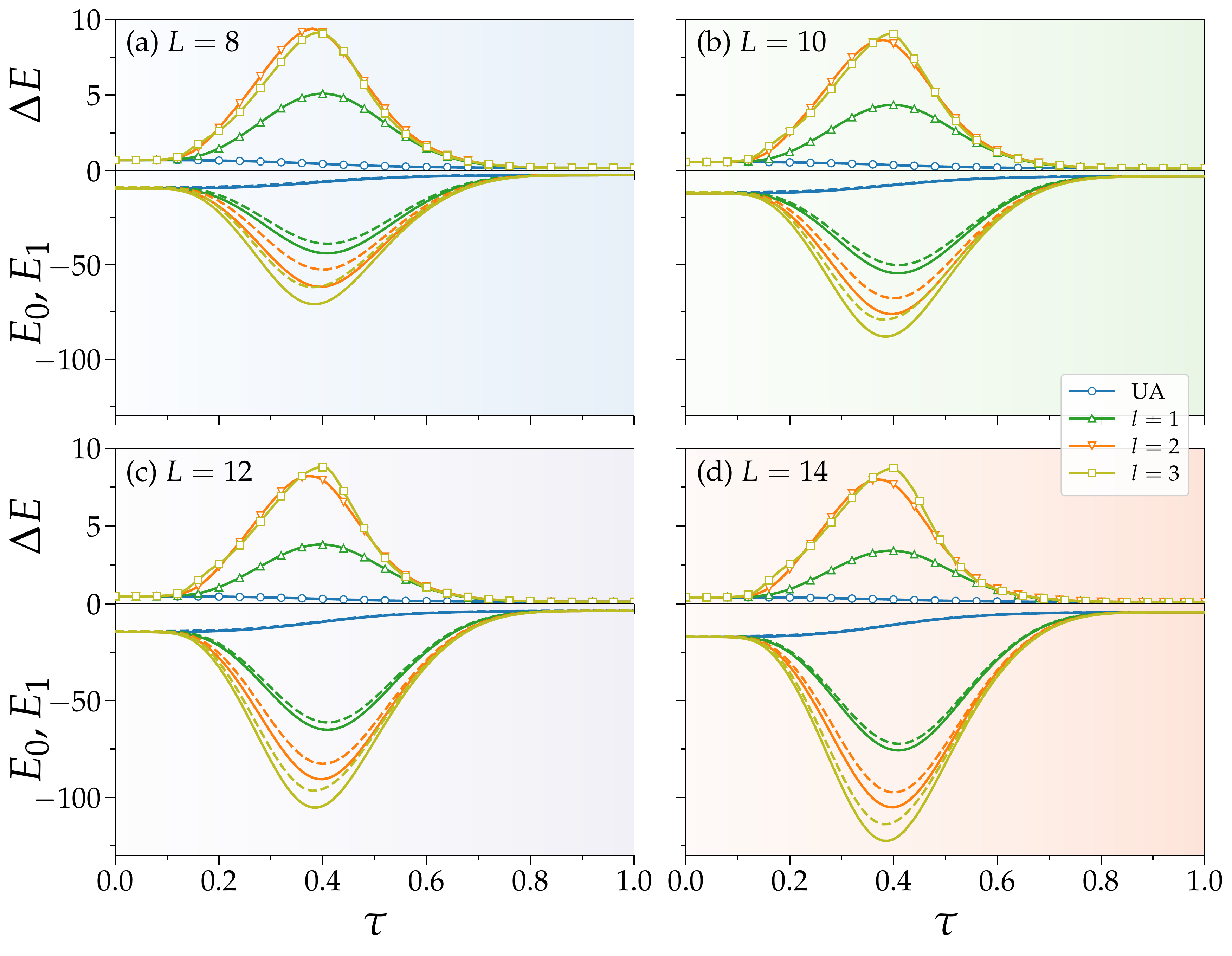}
\caption{
The time evolution of
the lowest two eigenvalues $E_0$ and $E_1$,
indicated by solid and dashed lines, respectively,
in the lower panels
and
the corresponding spectrum gap $\Delta E$
in the upper panels for the UA and CD ($l=1,2,3$) models
on the 1D chains of
(a) $L = 8$,
(b) $L = 10$,
(c) $L = 12$, and
(d) $L = 14$ sites at half filling.
The remaining parameters are
the same as in Fig.~\ref{fig:fidelity}.
}
\label{fig:gap}
\end{figure*}

Figure~\ref{fig:gap} monitors the lowest two eigenvalues, $E_0$ and $E_1$, and the
associated spectrum gap $\Delta E$ during the evolution for $L = 8,10,12$, and $14$.
In the UA model, the two eigenvalues monotonically increase
with the evolution time, as shown by
the solid and dashed blue lines in the lower panels of Figs.~\ref{fig:gap}(a)-\ref{fig:gap}(d),
and the spectrum gap monotonically decreases, as
shown in the upper panels.
Hence, the maximal spectrum gap appears at the beginning of the evolution.
In the CD models, the additional AGP terms first decrease the two eigenvalues
to reach the minimal values and then retrieve the energies in the UA model at $t=T$,
thus resulting in a valley shaped time-evolution of the spectrum.
Importantly, the two eigenvalues decrease differently in slopes,
i.e., the ground-state energy $E_0$ decreases steeper than the first-excited energy $E_1$,
which yields an increase in the spectrum gap $\Delta E$.
The gap maxima appear near the bottom of the spectrum valley.
In the case of $L = 14$, the maximal gap in the UA model is 0.4181,
while it increases to 3.4122, 7.9891, and 8.7280 in the {first-}, {second-}, and third-order CD protocols, respectively.
Nevertheless, we should note
that in the case of $L=8$ shown in the upper panel of Fig.~\ref{fig:gap}(a),
the third-order CD protocol
has the maximal spectrum gap $\Delta E=9.1050$ that is smaller than
$\Delta E=9.3783$ of the second-order CD protocol, despite the fact
that the better fidelity has been achieved for the third-order CD protocol
throughout the evolution in Fig.~\ref{fig:fidelity}(a).

\begin{figure*}
\includegraphics[width=\hsize]{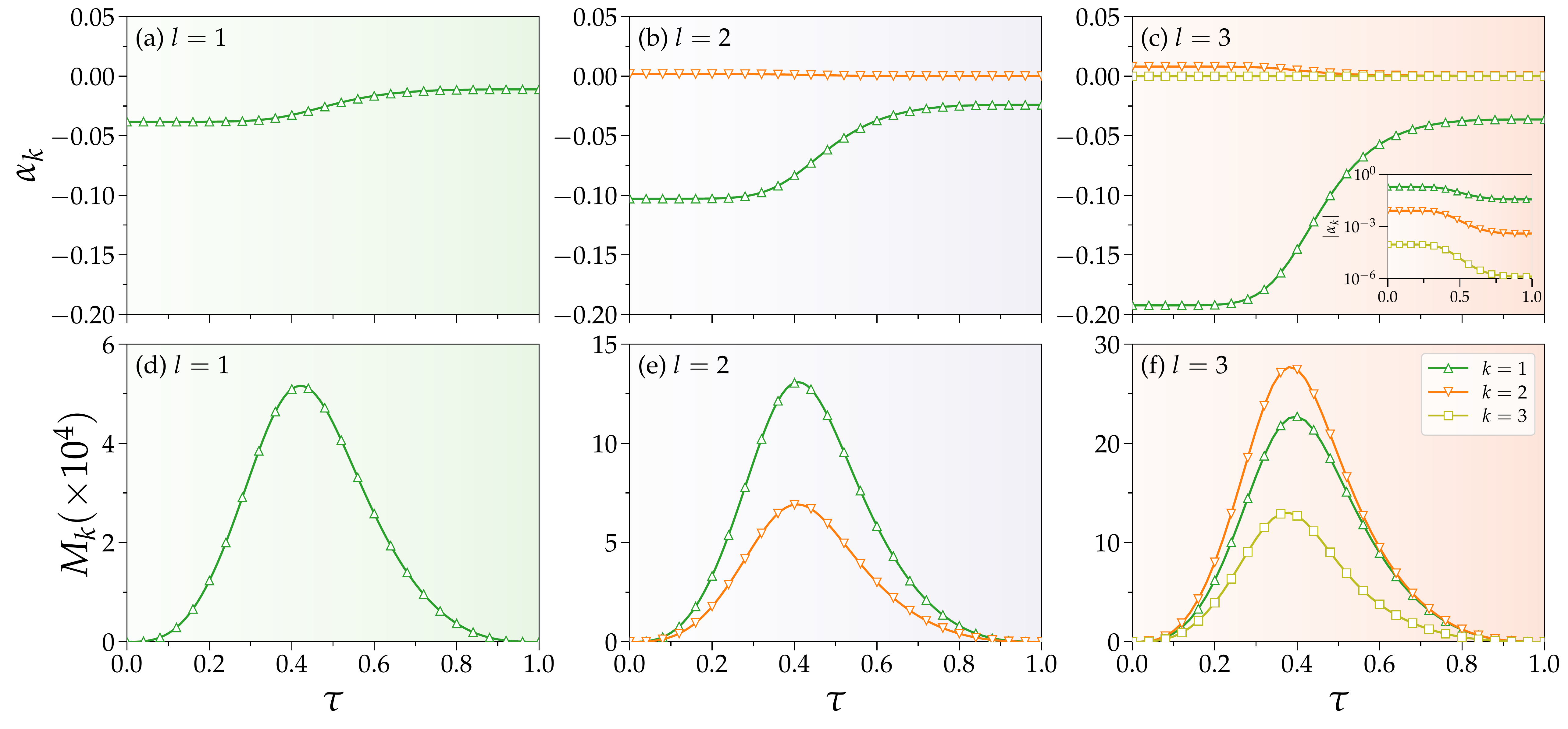}
\caption{
The time evolution of
(a-c) the optimal variational parameters $\alpha_k$ and
(d-f) the magnitudes $M_k$ of the corresponding AGP terms
for the CD models with
(a,d) $l=1$,
(b,e) $l=2$, and
(c,f) $l=3$.
The results are obtained for $L=14$ at half filling.
The remaining parameters are
the same as in Fig.~\ref{fig:fidelity}.
The inset in (c) is a semi-logarithmic plot of $|\alpha_k|$,
which demonstrates the exponential decrease of $|\alpha_k|$ in $k$.
}
\label{fig:alpha}
\end{figure*}

Figure~\ref{fig:alpha} shows the time evolution of the optimal
variational parameters $\alpha_k$
and the magnitudes $M_k$ of the AGP terms
for the CD ($l = 1,2,3$) models on $L = 14$ sites.
First, we note
that $\alpha_1$ is always negative for $l=1$,
as $\alpha_1=-\Gamma_1/\Gamma_2<0$ [see Eq.~(\ref{eq:linear2})]
is satisfied regardless of the Hamiltonian,
and Fig.~\ref{fig:alpha}(a) indeed numerically confirms this behavior.
Second, as discussed in Sec.~\ref{sec:remarks},
we find that for a given order $l$ of the CD protocol,
the absolute value $|\alpha_k|$ of the optimal variational parameter
becomes exponentially smaller with increasing $k$
[see, e.g., the inset of Fig.~\ref{fig:alpha}(c)].
Third, likewise,
we observe that $\Gamma_k$ increases exponentially in $k$.
For example, as shown in Fig.~\ref{fig:alpha}(f),
$M_k$ exhibits the maxima near $\tau = 0.4$ for $l=3$,
at which we find that
$(|\alpha_{1}|,|\alpha_{2}|,|\alpha_{3}|) \sim
(
1\times 10^{-1},
5\times 10^{-3},
5\times 10^{-5})$ and
$(\Gamma_{1},\Gamma_{3},\Gamma_{5})\sim
(
5\times 10^{9},
7\times 10^{12},
2\times 10^{16})$.
As a result, the magnitudes
$M_k$
for different $k$ values are in the same order,
indicating that the higher-order terms in the approximated
AGP also have significant contributions to the CD driving
even though $|\alpha_k|$ is exponentially small for large $k$.

\subsection{Filling dependence}

\begin{figure*}
\includegraphics[width=\hsize]{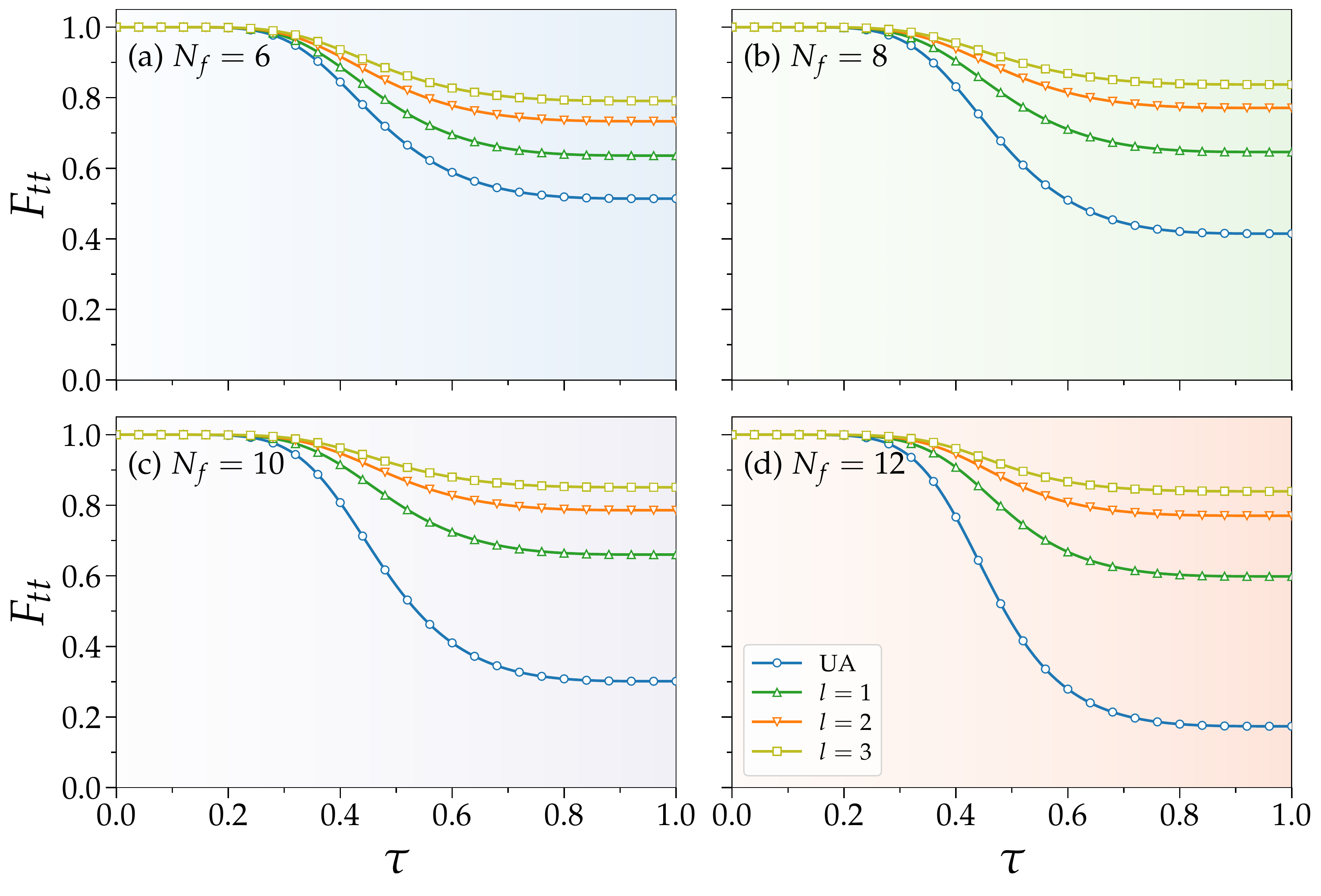}
\caption{
The time evolution of fidelity $F_{tt}$
for the UA model and the CD ($l=1,2,3$) models
on the 1D chain of $L=14$
sites at different fermion fillings:
(a) $N_{f}=6$,
(b) $N_{f}=8$,
(c) $N_{f}=10$, and
(d) $N_{f}=12$.
The results are obtained for the $S_z = 0$ sector
where the numbers of up and down fermions are equal,
i.e., $N_\uparrow = N_\downarrow = N_{f}/2$.
The remaining parameters are
the same as in Fig.~\ref{fig:fidelity}.
}
\label{fig:doping}
\end{figure*}

Figure~\ref{fig:doping}
shows the time evolution of fidelity $F_{tt}$ for the UA and CD ($l=1,2,3$) models
with $U=8$ on $L=14$ sites occupied by different numbers of fermions,
$N_{f} = 6,8,10$ and $12$, which correspond to
the Hilbert-space dimensions
$D=132496,
1002001,
4008004$,
and $9018009$, respectively, assuming that $N_\uparrow=N_\downarrow$.
Similarly to the cases at half filling, we find that
the higher order $l$ of the CD model yields the better fidelity,
which also demonstrates
the universal usefulness of the CD protocol
to the fermionic Hubbard model.
In the UA model, we find that
the final fidelity $F_{TT}$ decreases with increasing $N_{f}$ toward half filling.
This is expected because the initial state
$|\psi(0)\rangle=|n (0)\rangle$, which
is the ground state of $\hat{H}_J$, better approximates
the less-correlated target state.
Remarkably, however,
the improvement of the fidelity with increasing $l$ is most significant
for $N_{f}=12$, despite the largest dimension of the Hilbert space.
For example,
while the final fidelities for $N_{f}=6$
are $51.39\%$, $63.59\%$, $73.32\%$, and $79.09\%$ for the UA model and the CD models with $l=1$, 2, and 3, respectively,
those for $N_{f}=12$
are $17.37\%$, $59.85\%$, $77.03\%$, and $83.94\%$, respectively, indicating that the
$N_{f}=12$ cases achieve even better final fidelities than the $N_{f}=6$ cases for $l=2$ and $3$.
These results suggest that, while the fidelity improves systematically in the order $l$ of the CD model,
irrespectively of $N_{f}$ or $D$, the effectiveness of the CD driving depends
non-monotonically on $N_{f}$ and $D$.

\subsection{ $T$ and $U$ dependence}

\begin{figure}
\includegraphics[width=\columnwidth]{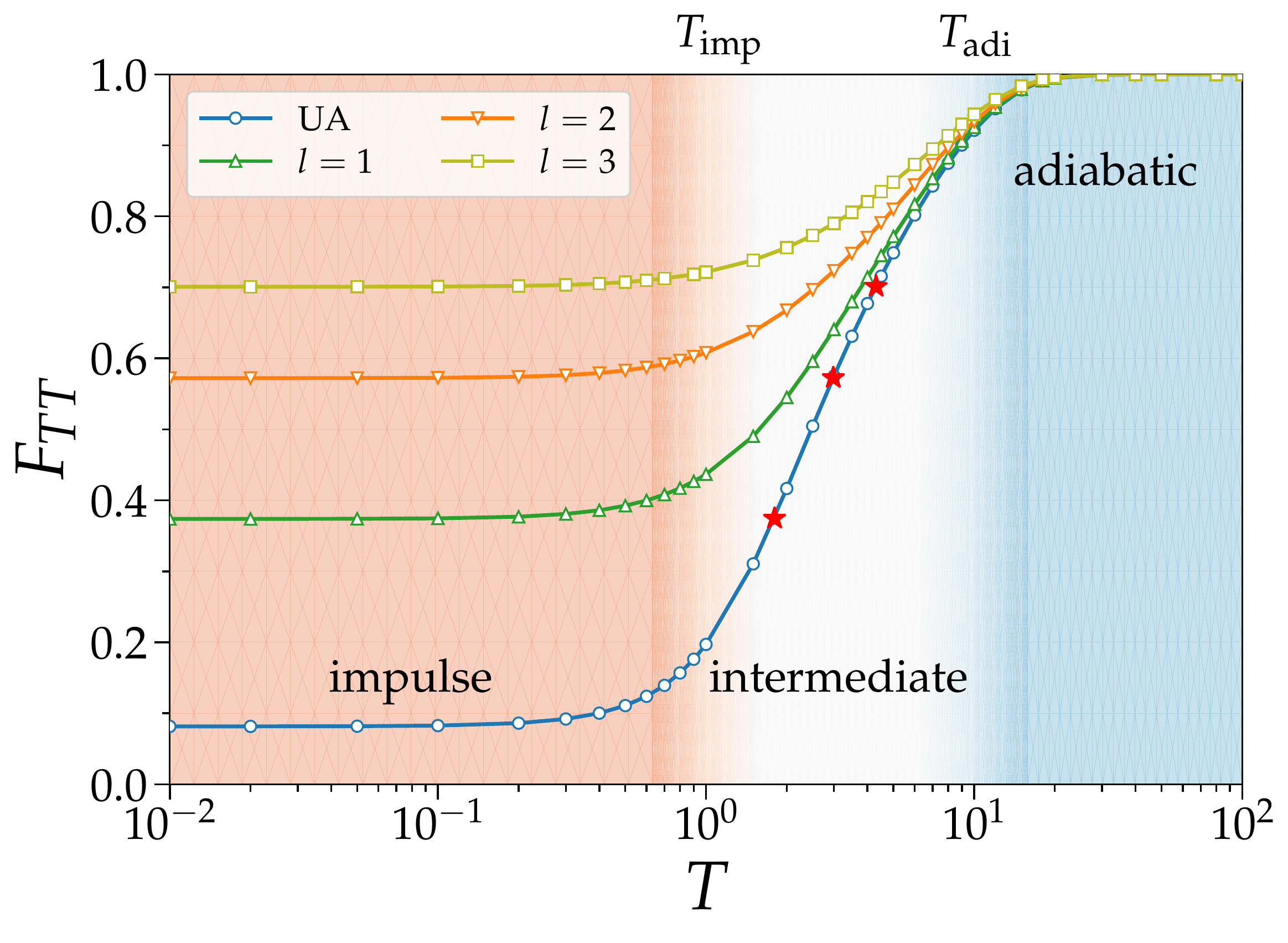}
\caption{
The final fidelity $F_{TT}$ as a function of the driving period $T$
for the UA and CD ($l=1,2,3$) models with $U=8$ on $L = 12$ sites at half filling.
For various $T$, the time step is fixed to be $\Delta t=0.001$.
The three red stars on the curve for the UA model
indicate the driving periods $T$ where the
$F_{TT}$ values are equal to $F_{TT}$ at $T=0.1$ for the CD models with
$l=1,2,$ and $3$.
The three regimes, i.e., impulse, intermediate, and adiabatic regimes,
distinguished by three different colors
crossover between themselves around $T_{\text{imp}}$
and $T_{\text{adi}}$ indicated at the top of the figure.
}
\label{fig:tau}
\end{figure}

\begin{figure}
\includegraphics[width=\columnwidth]{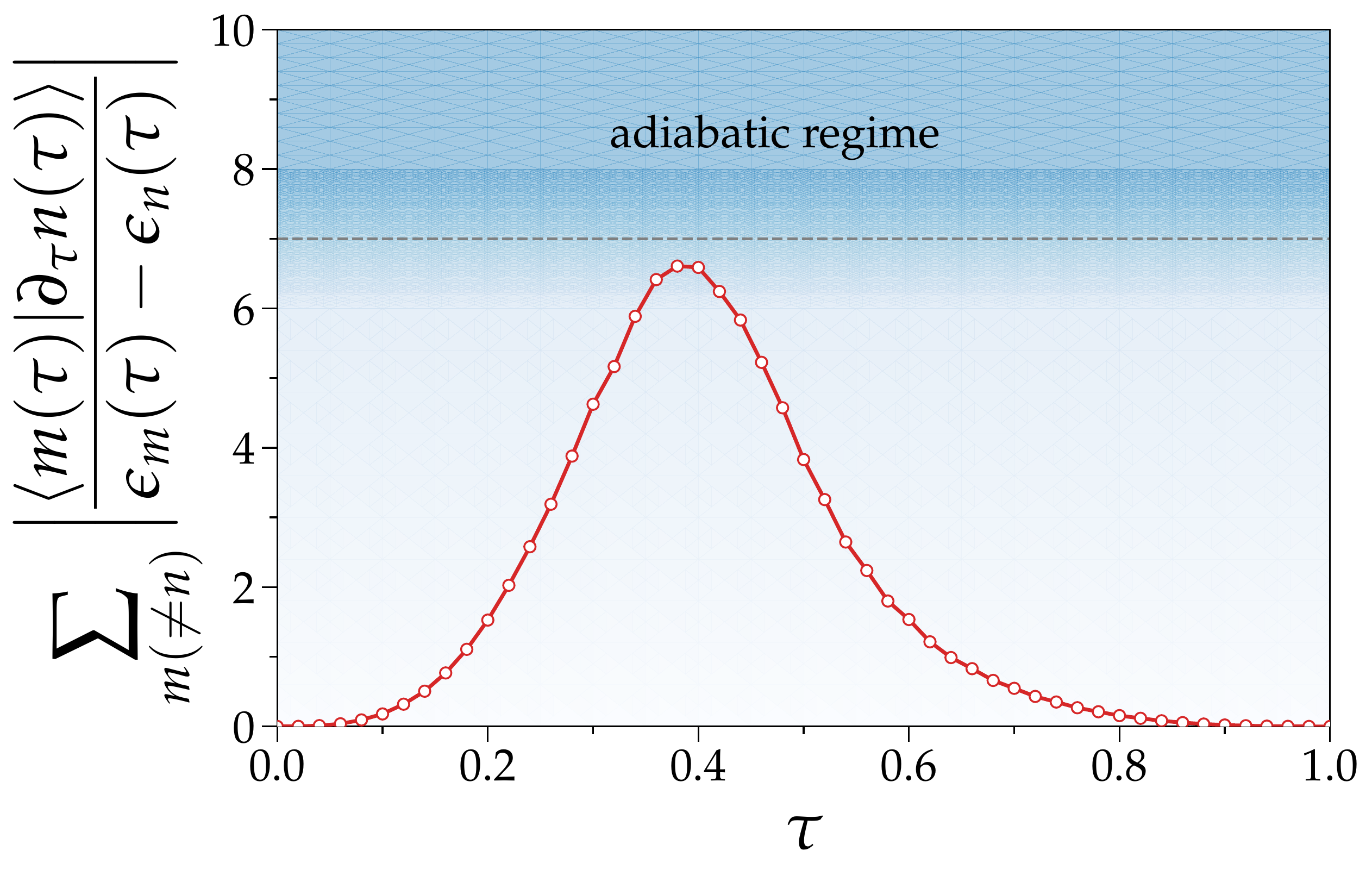}
\caption{
The time evolution of the quantity determining the adiabatic condition given
in Eq.~(\ref{eq:crit}) for the UA model with $U=8$ on $L = 12$ sites at half filling.
This is evaluated by using the Lanczos-based method
described in Ref.~\cite{SM}.
The dashed horizontal line is a guide for the eye.
}
\label{fig:crit}
\end{figure}

Figure~\ref{fig:tau} illustrates the final fidelity $F_{TT}$
as a function of the driving period $T$ for the UA and CD models with $U=8$ on $L=12$ sites at half filling.
All the results indeed show a monotonic behavior
where the smaller (larger) driving period $T$, i.e., faster (slower) driving,
yields the worse (better) fidelity.
Interestingly, we observe three distinct regimes:
an adiabatic regime for $T \gtrsim T_{\text{adi}} \sim 10$,
an impulse regime for $T \lesssim T_{\text{imp}} \sim 1.0$,
and
an intermediate regime for
$ T_{\text{imp}} \lesssim T \lesssim T_{\text{adi}}$
~\cite{NewJPhys.12.093025}.
In the adiabatic regime,
the time-evolved state accomplish almost perfect fidelity
even for the UA model.
It is well-known that the necessary condition for adiabaticity of quantum dynamics is given by
\begin{align}
\label{eq:eq43}
\sum_{m\,(\ne n)}\left| \frac{ \langle m(t) | \partial_t n(t) \rangle }{\epsilon_m(t) - \epsilon_n(t)} \right| \ll 1
\end{align}
for $t\in[0,T]$, assuming that the initial state $|\psi(0)\rangle$ is in the $n$th instantaneous
eigenstate $|n(0)\rangle$ of the initial Hamiltonian~\cite{PhysRevLett.58.1593,JMathPhys.49.125210,PhysRevLett.104.120401}.
This implies that the characteristic time $T_{\text{adi}}$ in our setting
is given by
\begin{align}
\label{eq:crit}
T_{\rm adi} \approx \max_{\tau\in[0,1]}
\left[ \sum_{m\, (\ne n) }
\left| \frac{ \langle m(\tau) | \partial_{\tau} n(\tau) \rangle }
{\epsilon_m(\tau) - \epsilon_n(\tau)} \right|
\right]
\end{align}
with $n = 0$, i.e., the instantaneous ground state of the Hamiltonian,
and adiabaticity is fulfilled when $T \gtrsim T_{\text{adi}}$.
Indeed, we find in Fig.~\ref{fig:crit} that $T_{\text{adi}}$
estimated from this criterion for the UA model is
in good accordance with the crossover boundary between the intermediate and adiabatic regions shown in Fig.~\ref{fig:tau}
(see Figs.~\ref{fig:fidelity5} and \ref{fig:TadiU} in Ref.~\cite{SM} for the results with different values of $U$).

On the other hand, in the impulse regime found in Fig.~\ref{fig:tau}, the
system is essentially frozen to stay in the initial state for the UA model [see Fig.~\ref{fig:fidelity3}(a)].
This is reasonable because in this region the driving is so fast that
the system has little time to react.
The characteristic time scale $T_{\text{imp}}$ below which the quantum state $|\psi(t)\rangle$ cannot follow the dynamics is determined
approximately by the inverse of the spectrum gap $\Delta E$~($= 0.4822$ for $L=12$) at $t=0$,
which is as large as $\sim2.1$ for our case studied here (see Fig.~\ref{fig:fidelity5} in Ref.~\cite{SM}
for the $U$ dependence).
Moreover, it is highly intriguing
to find that even in the fast driving regime $T \lesssim T_{\text{imp}}$,
the higher order CD models can achieve better fidelity $F_{TT}$,
suggesting that an extremely fast CD driving with $T\to0^+$ is
in principle possible without much deteriorating the fidelity,
provided that the order $l$ of the CD models is large enough
(see Fig.~\ref{fig:fidelity4} in Ref.~\cite{SM}).
We also note that
in the intermediate regime, the final fidelity
$F_{TT}$ increases almost logarithmically with $T$ for the UA model as well as the CD models.

To better understand how the CD protocol can speed up the driving process,
we compare the fidelity $F_{TT}$ of the CD model
with that of the UA model.
Specifically, at $T = 0.1$,
the fidelity in the $l=1$ protocol is 37.46\%.
On the $F_{TT}-T$ curve for the UA model in Fig.~\ref{fig:tau}, this fidelity
can be reached for a much slower driving with $T = 1.8$ (indicated by a red star in Fig.~\ref{fig:tau}).
In other words, the $l=1$ CD protocol realizes
an 18 times speedup.
Accordingly,
the $l=2$ and $3$ CD protocols realize
30 and 43 times speedup, respectively, against the UA model
to reach the corresponding fidelities $F_{TT}$ at $T=0.1$.
Although the speedup depends on the driving period $T$,
generally speaking,
the CD protocol
is more effective for the impulse regime
where the UA model has a very low fidelity
and the first oder CD driving
already can significantly boost the fidelity.
On the other hand, in the adiabatic regime,
the fidelity $F_{TT}$ for the UA model is already close to unity, and hence
the CD protocol is not necessary.

\begin{figure}
\includegraphics[width=\columnwidth]{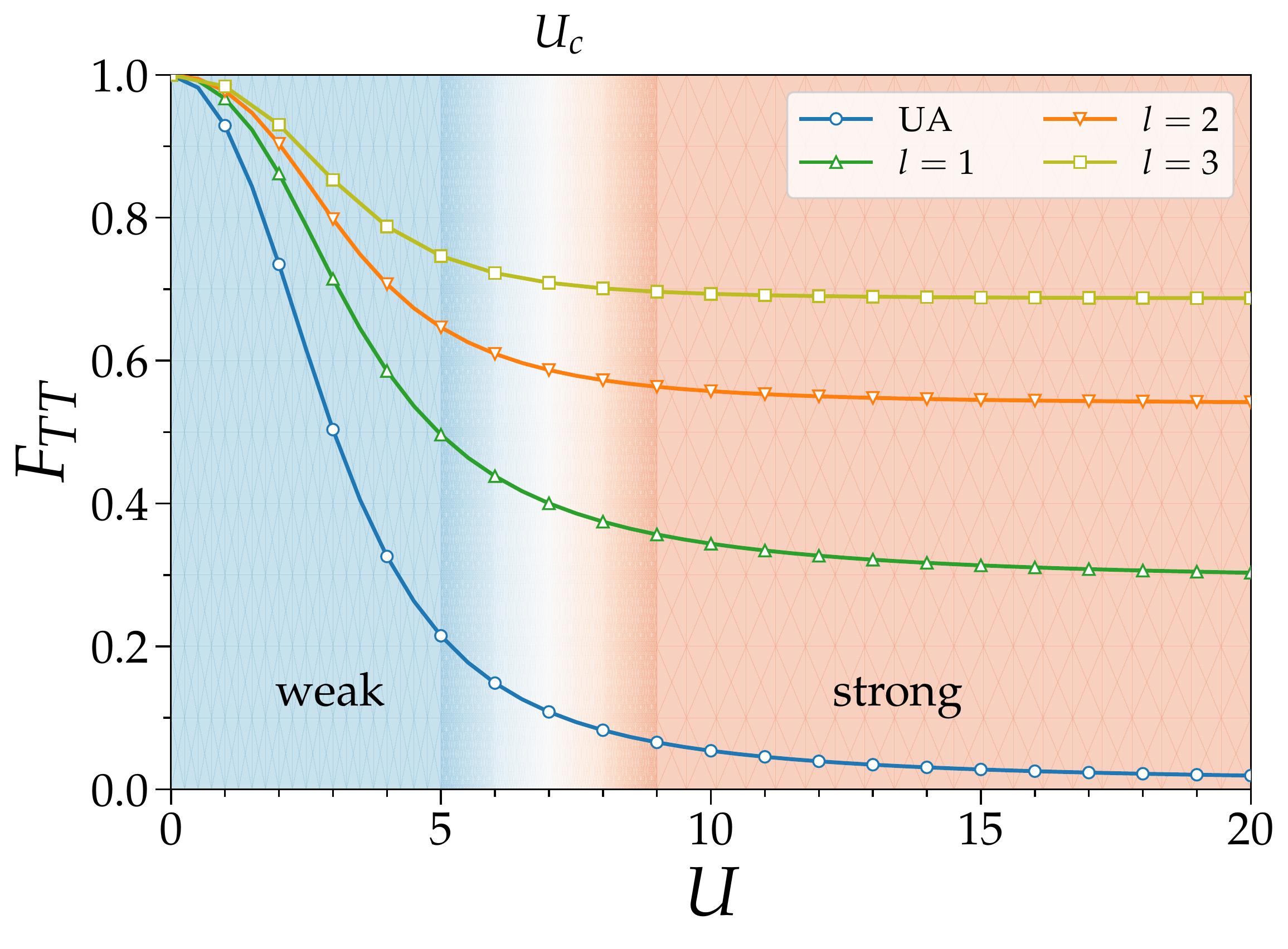}
\caption{
The final fidelity $F_{TT}$ as a function of the interaction strength
$U$ for the UA and CD ($l=1,2,3$) models on $L = 12$ sites at half filling.
The remaining parameters are the same as in Fig.~\ref{fig:fidelity}.
Weak- and strong-correlation regimes crossover around $U_c$
indicated at the top of the figure.
}
\label{fig:U}
\end{figure}

Figure~\ref{fig:U} shows the final fidelity
$F_{TT}$ as a function of
the interaction strength $U$ for the UA and CD models on $L=12$ sites
with the driving period $T = 0.1$.
At $U \ll 1$,
where $\hat{H}_U$ can be considered as a perturbation to $\hat{H}_J$,
the fidelity between the initial state
$|\psi(0)\rangle=|n (0)\rangle$ and
the target state $|n(T) \rangle$, i.e., the ground state of the Hubbard model $\hat{H}_{\rm HB}$,
is already large, and hence
the high fidelity can be achieved even for the UA model.
On the other hand, as shown in Fig.~\ref{fig:U}, for $U \gg 1$, $F_{TT}$ dramatically decreases as expected.
This implies that,
in our setting of the UA and CD models in Eqs.~(\ref{eq:hubbard}) and (\ref{eq:hcd}), respectively,
the ground state of the Hubbard model with large $U$
is much harder to prepare.
For all the UA and CD protocols shown in Fig.~\ref{fig:U}, we also observe that there exist approximately
two distinct regimes:
a weak-correlation regime for $U \lesssim U_c \sim 7$ and
a strong-correlation regime for $U \gtrsim U_c$.
The CD protocol is more effective for the strong-correlation regime
in the sense that it increases the final fidelity
more significantly as compared with the UA  protocol for the same $U$.

\section{Conclusions}
\label{sec:conclusion}

In summary, we have applied the variational CD driving protocol
proposed in
Ref.~\cite{PhysRevLett.123.090602}
to the 1D two-component fermionic Hubbard model.
We have formalized the variational optimization procedure of the CD driving
and shown that the optimal variational parameters are obtained deterministically by solving
a set of linear equations whose coefficients are given by
the squared Frobenius norms of the nested commutators $\hat{O}_k$.
We have also devised an algorithm
to construct analytical expressions of the nested commutators $\hat{O}_k$,
which enables us to simulate systems up to
$L=14$ sites with driving order $l \leqslant 3$.
We have shown
that the fidelity $F_{tt}$ dramatically increases
with increasing $l$ throughout the evolution.
Moreover, we have found that
the CD driving protocol
is more effective for
the fast driving with the smaller driving period $T$ and
the strong-correlation regime with the larger interaction strength $U$,
where the increase of the final fidelity $F_{TT}$
is most significant
when it is compared with the UA driving protocol.

Our results
demonstrate the usefulness of the variational CD protocol
for interacting fermions,
and would be beneficial for further exploring fast ground-state preparation
protocols for many-body fermionic systems, not only on classical
computers but also on quantum devices in the foreseeable future.
Indeed, establishing a variational-CD-inspired ansatz
for the ground-state preparation of many-body systems on a quantum computer
is an interesting issue to be addressed.
A possible route for this is to combine
the discretized quantum adiabatic process~\cite{Shirakawa2021} with
an efficient implementation of unitary operators generated by
higher-order nested commutators~\cite{PhysRevResearch.4.013191}.

\acknowledgements
Part of the numerical calculations have been performed
using the HOKUSAI BigWaterfall system at RIKEN (Project IDs: Q22551 and Q22525).
This work is supported by Grant-in-Aid for Research Activity start-up (No.~JP19K23433),
Grant-in-Aid for Scientific Research (C) (No.~JP22K03520),
Grant-in-Aid for Scientific Research (B) (No.~JP18H01183), and
Grant-in-Aid for Scientific Research (A) (No.~JP21H04446) from MEXT, Japan.
This work is also supported in part by the COE research grant in computational science from
Hyogo Prefecture and Kobe City through Foundation for Computational Science.
A Fortran package that generates the data reported in this paper is available at~\cite{SourceCode},
which have used the Bessel function subroutine~\cite{BesselFunction}
by John Burkardt under the GNU LGPL licence.

\appendix

\section{ Proof of Eq.~(\ref{eq:OO}) } \label{sec:proof}
\label{appendix:a}
\numberwithin{equation}{section}

From Eq.~(\ref{eq:Ok2}), it is easy to verify
\begin{align}
\label{eq:eq39}
\text{Tr}\left(\hat{O}_m \hat{O}_n \right)
=  - \text{Tr}\left(\hat{O}_{m+1} \hat{O}_{n-1} \right).
\end{align}
This relation holds for $m \ge 0$ and $n \ge 1$, implying that
$\text{Tr}\left(\hat{O}_m \hat{O}_n \right)=0$ when $n = m + 1$.
Then, using Eq.~(\ref{eq:Ok_dag}), one can readily show that
\begin{alignat}{1}
\label{eq:eq40}
\left\langle \hat{O}_{2m}, \hat{O}_{2k}\right\rangle_{\text{F}}
& = \text{Tr}\left(\hat{O}_{2m}^\dag \hat{O}_{2k}\right) \notag \\
& = (-1)^{m+k}\text{Tr}\left(\hat{O}_{m+k}^2\right) \notag \\
& = \text{Tr}\left(\hat{O}_{m+k}^\dag \hat{O}_{m+k}\right) \notag \\
& = \left|\left|\hat{O}_{m+k}\right|\right|_{\text{F}}^2,
\end{alignat}
which thus proves Eq.~(\ref{eq:OO}).




\renewcommand{\thetable}{S\arabic{table}}
\renewcommand{\thefigure}{S\arabic{figure}}
\renewcommand{\thetable}{S\arabic{table}}
\renewcommand\theequation{S\arabic{equation}}
\setcounter{table}{0}
\setcounter{figure}{0}
\setcounter{equation}{0}
\setcounter{enumi}{0} 
\renewcommand{\bibnumfmt}[1]{[S#1]}
\renewcommand{\citenumfont}[1]{S#1}

\onecolumngrid

\clearpage

\begin{center}
{\bf \large
\textit{Supplemental Material}:\\
Variational counterdiabatic driving of the Hubbard model for ground-state preparation
}

\vspace{0.2 cm}

\vspace{0.2 cm}

Q.~Xie,$^{1,2}$ Kazuhiro~Seki,$^{1}$
and Seiji Yunoki,$^{1,2,3,4}$

\vspace{0.1 cm}

{\small
{\it
$^1$Quantum Computational Science Research Team, RIKEN Center for Quantum Computing (RQC), Saitama 351-0198, Japan

$^2$Computational Condensed Matter Physics Laboratory, RIKEN Cluster for Pioneering Research (CPR), Saitama 351-0198, Japan

$^3$Computational Materials Science Research Team, RIKEN Center for Computational Science (R-CCS),  Hyogo 650-0047,  Japan

$^4$Computational Quantum Matter Research Team, RIKEN Center for Emergent Matter Science (CEMS), Saitama 351-0198, Japan
}
}

\end{center}

\subsection*{Driving function}

\begin{figure}[h]
\includegraphics[width=.5\columnwidth]{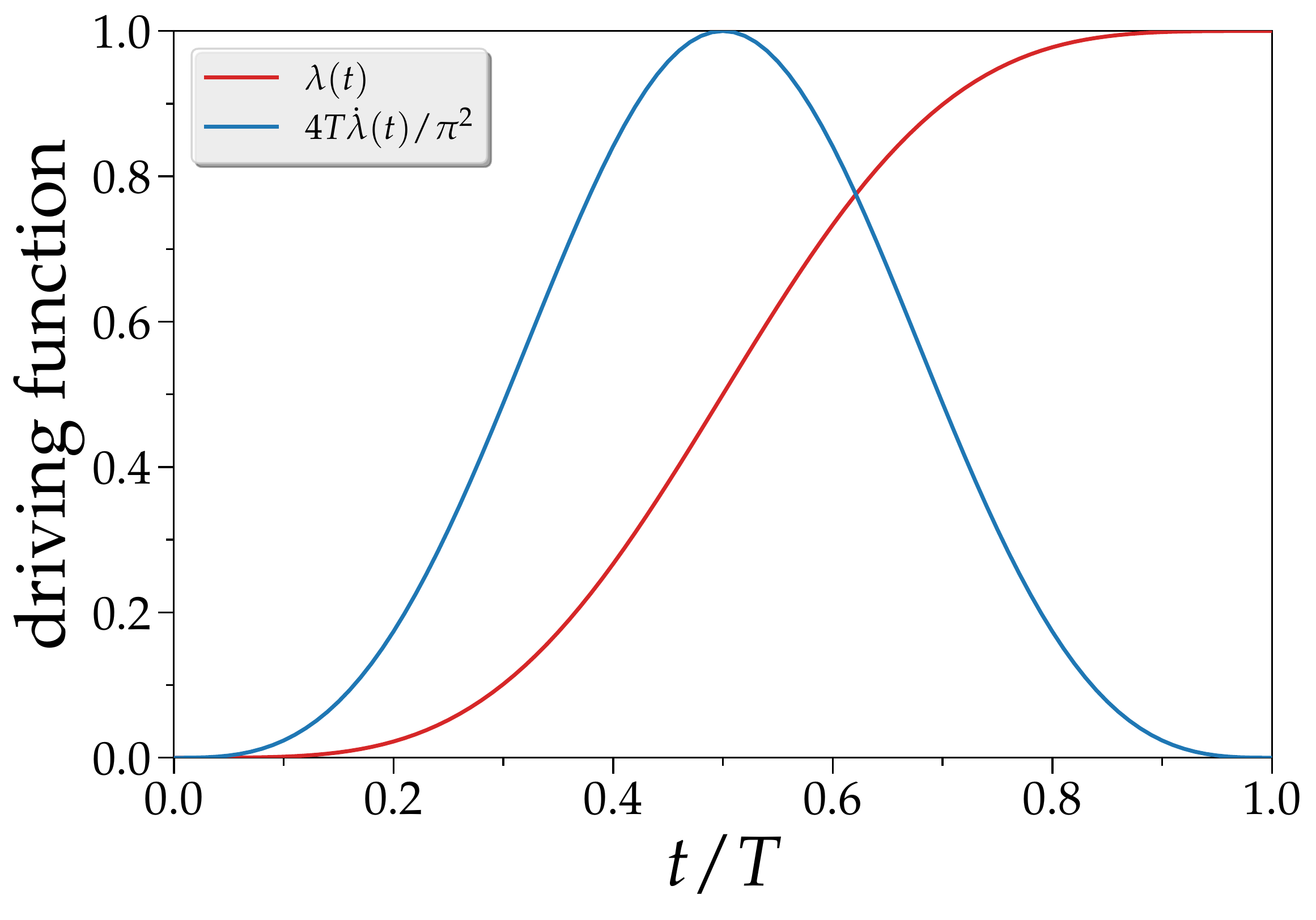}
\caption{ Driving function $\lambda(t)$ and its time derivative $\dot{\lambda}(t)$.
}
\label{fig:lambda}
\end{figure}

Figure~\ref{fig:lambda} shows the driving function $\lambda(t)$ defined in Eq.~(\ref{eq:lambda}) in the main text
and its time derivative $\dot{\lambda}(t)$ in Eq.~(\ref{eq:lambda_t}).

\subsection*{ Direct approach }

For small systems, we can take the direct approach,
in which the $\hat{O}_k$ operators defined in Eq.~(\ref{eq:Ok}) are treated by
direct matrix-matrix multiplications recursively via Eq.~(\ref{eq:Ok2}).
We use this approach as a benchmark to test
the constructive approach described below.

\subsection*{Convergence and systematic error}

\begin{figure}
\includegraphics[width=.8\columnwidth]{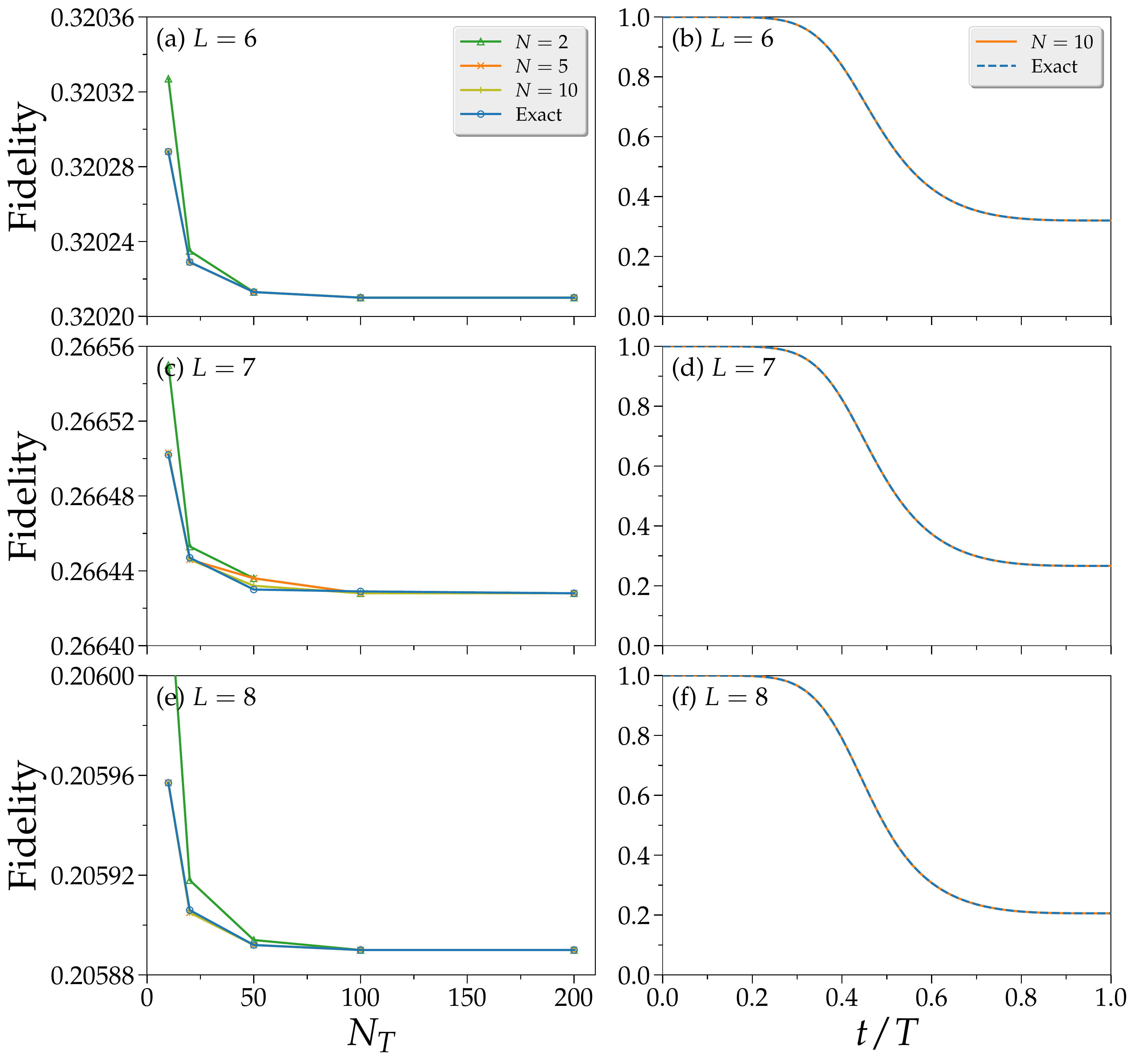}
\caption{
Comparison of the results obtained by the full diagonalization method
and the Chebyshev polynomial expansion method
for the UA model with $U=8$ on different system sizes
$L = 6,7$, and $8$ at half filling.
Here the evolution period is set to be $T=0.1$.
Left panels (a, c, e) show the final fidelity $F_{TT}$ as a function of the number $N_T$ of time steps.
In the Chebyshev polynomial expansion method, three different expansion orders ($N=2,5$, and $10$) are used.
Note that since the evolution period $T$ is fixed,
the time interval $\Delta t=T/N_T$ becomes smaller with increasing
$N_T$.
Right panels (b, d, f) show the time evolution of fidelity $F_{tt}$.
We set $N = 10$ in the Chebyshev polynomial expansion method and $\Delta t=0.001$.
}
\label{fig:check}
\end{figure}

Figure~\ref{fig:check}
compares the results of the fidelity for the UA model obtained by the full diagonalization method
and the Chebyshev polynomial expansion method for small systems $L = 6,7$, and $8$.
Due to the small time interval $\Delta t = T/N_T = 0.001$,
a small Chebyshev expansion order $N = 10$ is enough to obtain the converged results to the exact values.
Note also that the systematic error due to the discretization of the time in the time-evolution operator
[see Eq.~(\ref{eq:psi_t}) in the main text] is also
negligible when $\Delta t$ is as small as 0.001, as shown in the left panels of Fig.~\ref{fig:check}.

\subsection*{Numerical results for small systems}

\begin{figure}
\includegraphics[width=.8\columnwidth]{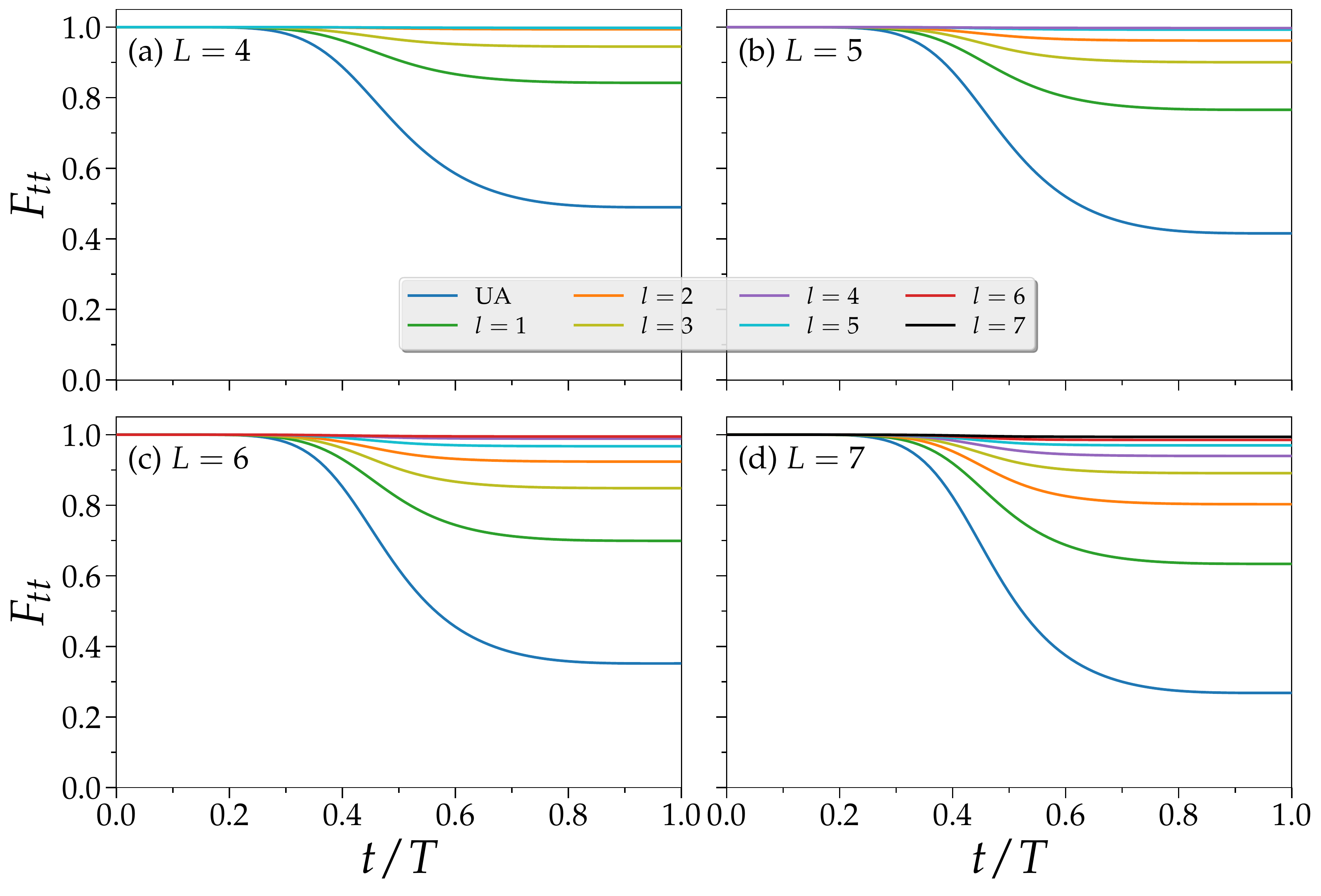}
\caption{
The time evolution of fidelity $F_{tt}$ for the UA model and the CD models with different orders $l$ on the 1D chains of
(a) $L = 4$,
(b) $L = 5$,
(c) $L = 6$, and
(d) $L=7$ sites at half filling.
In (b) and (d), $N_\uparrow=N_\downarrow+1$ and thus the total $S_z=1/2$.
The remaining parameters are the same as in Fig.~\ref{fig:fidelity} in the main text.
The largest final fidelities $F_{TT}$ are $99.80\%$, $99.73\%$, $99.49\%$, and $99.39\%$
for $L= 4,5,6$, and $7$ with $l=4,5,6$, and $7$, respectively.
}
\label{fig:fidelity2}
\end{figure}

Figure~\ref{fig:fidelity2} shows the time evolution of fidelity $F_{tt}$ for the UA and CD models with $U=8$
on small sites ($L=4,5,6$, and $7$) at half filling, i.e., $N_{f} = L$,
with $N_{\uparrow} = N_{\downarrow}$ for even $N_{f}$
and $N_{\uparrow} = N_{\downarrow} + 1$ for odd $N_{f}$.
Here, we use the direct approach and set the driving period $T=0.1$.
The higher-order CD protocols can be applied for these small systems, where the final fidelity $F_{TT}$ as well as
$F_{tt}$ during the whole driving period approaches almost one, implying that essentially the perfect CD evolution is achieved.

\begin{figure}
\includegraphics[width=.8\columnwidth]{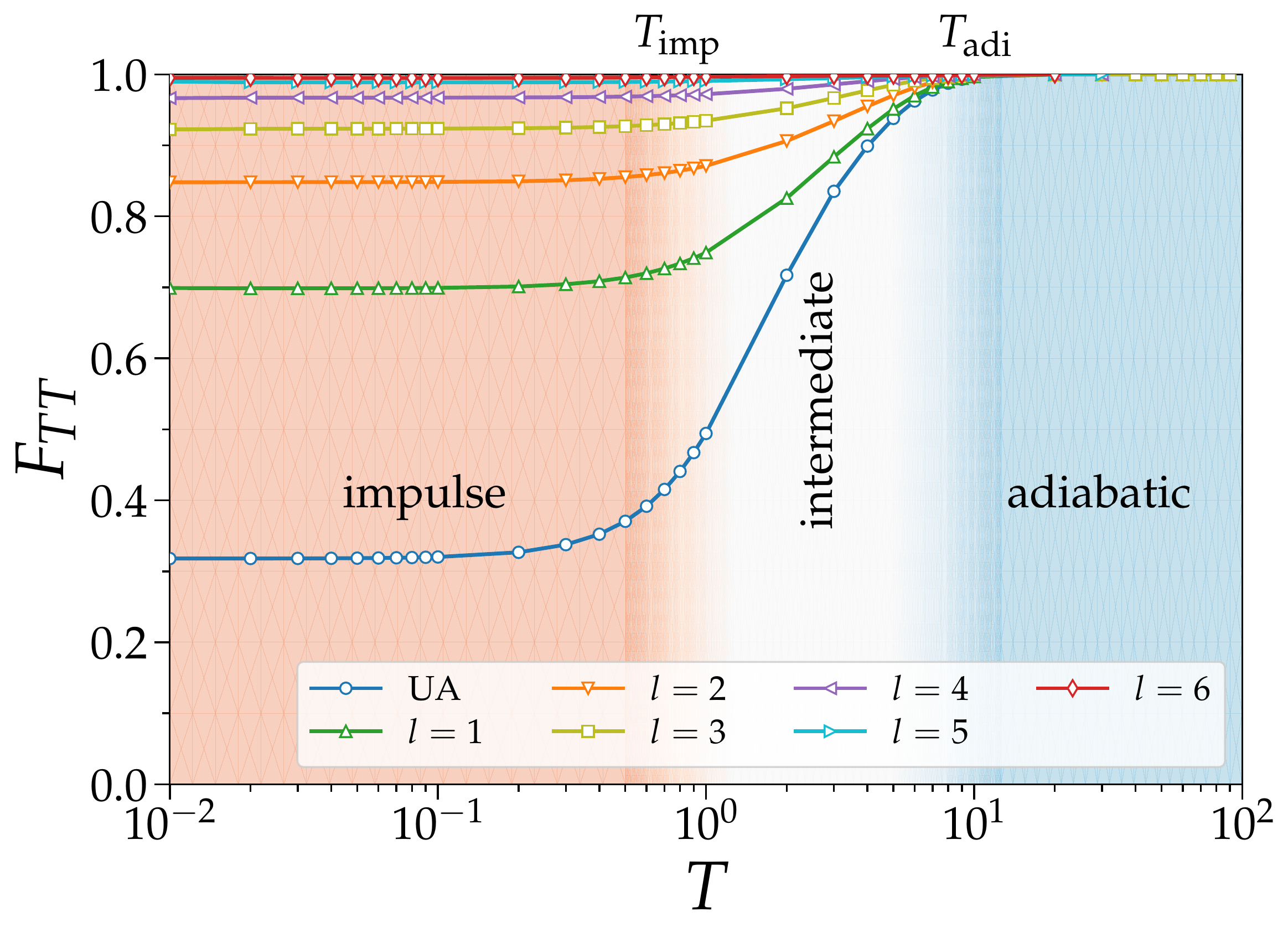}
\caption{
The final fidelity $F_{TT}$ as a function of the driving period $T$
for the UA and CD ($l=1,2,\cdots,6$) models with $U=8$ on $L = 6$ sites at half filling.
For various $T$, the time step is fixed to be $\Delta t=0.001$.
}
\label{fig:fidelity4}
\end{figure}

Figure~\ref{fig:fidelity4} shows the final fidelity $F_{TT}$ as a function of the driving period $T$
for the UA and CD ($l=1,2,\cdots,6$) models with $U=8$ on $L = 6$ sites at half filling calculated using the direct approach.
Although the system size used here is smaller, we find that the crossover boundaries among the impulse, intermediate, and adiabatic
regions are similar to those indicated in Fig.~\ref{fig:tau}, for which the system size $L=12$ is considered. The spectrum gap
$\Delta E$ for the UA model at $t=0$, which is the largest during the time evolution, is $0.8901$ for $L=6$.
The characteristic time $T_{\rm adi}$ defined in Eq.~(\ref{eq:crit}) is $\sim9$ for the UA model with $U=8$ on $L=6$ sites.
We should also emphasize in Fig.~\ref{fig:fidelity4} that the final fidelity $F_{TT}$ approaches to one with
increasing the order $l$ for the CD model even when $T\to0^+$.

\subsection*{ Constructive approach }

As discussed in Sec.~\ref{sec:const_cd}, the UA and CD models are composed of the summation
of multi-fermion-operator products of the form
\begin{align}
\hat{h}_{m, x}
= \lambda(t)^{m} \beta_x \hat{c}^\dag_i \hat{c}^\dag_j \cdots \hat{c}_p \hat{c}_q \cdots.
\end{align}
We refer to each of these products as a Hamiltonian term.
The total number $N_{\text{term}}$ of terms in the UA model $\hat{H}(t)$ scales linearly in the system size $L$.
For instance, under open-boundary conditions,
an $L$-site chain has $2(L-1)$ hopping terms for each spin and $L$ onsite interacting terms
(assuming nonzero $J$ and $U$).
Thus, $N_{\text{term}} = 5L-4\sim\mathcal{O}(L)$.

There are three different conventions for the order of fermion operators in Hamiltonian terms:
(1) general,
(2) normal-ordered (NO),
and (3) special normal-ordered (SNO) forms.
In the general form, the creation and annihilation operators
have no particular order.
In the NO form, all creation operators
are on the left of annihilation operators.
In the SNO form, the creation and annihilation operators
in a NO term
are both ordered according to the indexes, e.g.,
$i<j<\cdots$ and $p<q<\cdots$.

A NO term can be transformed into a SNO term
by simple permutations.
Except for a possible minus sign due to the anticommutative relation
(i.e., $\hat{c}^\dag_j\hat{c}^\dag_i = - \hat{c}^\dag_i \hat{c}^\dag_j$,
$\hat{c}_q\hat{c}_p = - \hat{c}_p \hat{c}_q$ ),
no additional term arises.
However, when one transforms a general term into the NO form,
many additional terms emerge in general
because $\hat{c}_i \hat{c}^\dag_j = \delta_{ij} - \hat{c}^\dag_j \hat{c}_i$.
Here we devise an algorithm to implement the transformation
among these three different forms of Hamiltonian terms,
and thus we are able to compute the commutator $\hat{O}_k = [\hat{H}, \hat{O}_{k-1}]$
in an analytic representation.

We now describe the constructive approach,
which has been implemented in our code available at~\cite{SourceCode_sm}.
Before the time-evolution calculation,

\begin{enumerate}

\item
We build the analytical representation of the UA model $\hat{H}(t)$.
The hopping part $\hat{H}_J$ has $m = 0$
and the interacting part $\lambda(t) \hat{H}_U$ has $m = 1$.

\item
The $\hat{O}_k$ operators for $k=1,2,\cdots,2l$ are constructed recursively
starting with $\hat{O}_0 = \hat{H}_U$.

\item
All the diagonal terms in $\hat{O}_k^\dag \hat{O}_k$ are constructed.

\item
The $S^{k,m}$ are calculated for $ 1 \leqslant k \leqslant 2l$ and $ 0 \leqslant m \leqslant 2l-1$
by tracing the diagonal terms in $\hat{O}_k^\dag \hat{O}_k$ over the many-body bases.

\item
The initial model $ \hat{H}(t=0) = \hat{H}_{\text{CD}}^{(l)}(t=0)$ is constructed and
the initial state $|\psi(t=0)\rangle = |n(t=0)\rangle$, i.e., the ground state of $\hat{H}(t=0)$,
is calculated.

\item
The ground state $|n(t=T)\rangle$ of $ \hat{H}(t=T) = \hat{H}_{\text{CD}}^{(l)}(t=T)$ at the final time $T$
is also calculated by using the Lanczos algorithm~\cite{Lanczos1950AnIM_sm}
for the evaluation of $F_{T t}$.

\end{enumerate}

The detailed procedure at each time step $t_i$ is as follows. Staring with $i=1$ and setting $t_{i=0}=0$ and $t_{i=1}=\Delta t$,
\begin{enumerate}
\setcounter{enumi}{6}
\item
The time-evolved state $|\psi(t_i)\rangle$ is calculated
through $  |\psi(t_i)\rangle = e^{ -i\Delta t \hat{H}[\lambda(t_{i-1})]} | \psi(t_{i-1})\rangle$
by using the Chebyshev polynomial expansion method.
The lowest two eigenvalues of $\hat{H}(t_i)$ are calculated
by using the block-Lanczos algorithm~\cite{GOLUB1977361_sm}, if necessary.
The same procedure is applied for the CD model $\hat{H}_{\rm CD}^{(l)}$ by replacing $\hat{H}$ with $\hat{H}_{\rm CD}^{(l)}$.
\item
The UA model $\hat{H}(t_i)$ is constructed.
\item
The ground state
$| n(t_i) \rangle$ of $\hat{H}(t_i)$ is calculated by using the Lanczos algorithm.
\item Fidelities $F_{tt}$, $F_{0t}$ and $F_{T t}$ are calculated. If $t_i=T$, exit.
\item
$\Gamma_k$ in Eq.~(\ref{eq:Gamma_k2}) are calculated for $1 \leqslant k \leqslant 2l$ by using $S^{k,m}$
already evaluated in step 4.
\item
The optimal parameters $\alpha_k$ ($k = 1,2,\ldots, l$)
are obtained by solving the set of the linear equations in Eq.~(\ref{eq:linear2}).
\item
An analytical form of $\hat{H}_{\text{CD}}^{(l)}(t_i)$
is constructed by adding the AGP terms
$ i \dot{\lambda} \alpha_k \hat{O}_{2k-1}$ with $k = 1,2,\ldots, l$ to $\hat{H}(t_i)$.
%
%
%
%
%
\item Go back to Step 7 with the next time step $t_i \rightarrow t_{i+1}=t_i+\Delta t$.
\end{enumerate}

As an example, we show the analytical forms of $\hat{H}[\lambda(T)=1] = \hat{H}_{\text{HB}}$
and $\hat{O}_k[\lambda(T) = 1]$ operators for $L = 2$ and $l = 1$
generated by our code available at~\cite{SourceCode_sm}.
The Hubbard model $\hat{H}_{\text{HB}}$ with $U=8$ is given by
\begin{verbatim}
6
c+ 1  c 2          0    -0.100000E+01   0.000000E+00
c+ 2  c 1          0    -0.100000E+01   0.000000E+00
c+ 3  c 4          0    -0.100000E+01   0.000000E+00
c+ 4  c 3          0    -0.100000E+01   0.000000E+00
c+ 1 c+ 3  c 1  c 3          1    -0.800000E+01  -0.000000E+00
c+ 2 c+ 4  c 2  c 4          1    -0.800000E+01  -0.000000E+00
\end{verbatim}
The $\hat{O}_k$ operators are given by \\
\\
$\hat{O}_0 = \hat{H}_U$:
\begin{verbatim}
2
c+ 1 c+ 3  c 1  c 3          0    -0.800000E+01  -0.000000E+00
c+ 2 c+ 4  c 2  c 4          0    -0.800000E+01  -0.000000E+00
\end{verbatim}
$\hat{O}_1 = [\hat{H}, \hat{O}_0]$:
\begin{verbatim}
8
c+ 1 c+ 3  c 1  c 4          0    -0.800000E+01  -0.000000E+00
c+ 1 c+ 3  c 2  c 3          0    -0.800000E+01  -0.000000E+00
c+ 1 c+ 4  c 1  c 3          0     0.800000E+01   0.000000E+00
c+ 1 c+ 4  c 2  c 4          0     0.800000E+01   0.000000E+00
c+ 2 c+ 3  c 1  c 3          0     0.800000E+01   0.000000E+00
c+ 2 c+ 3  c 2  c 4          0     0.800000E+01   0.000000E+00
c+ 2 c+ 4  c 1  c 4          0    -0.800000E+01  -0.000000E+00
c+ 2 c+ 4  c 2  c 3          0    -0.800000E+01  -0.000000E+00
\end{verbatim}
$\hat{O}_2 = [\hat{H}, \hat{O}_1]$:
\begin{verbatim}
20
c+ 1 c+ 3  c 1  c 3          0    -0.320000E+02   0.000000E+00
c+ 1 c+ 3  c 2  c 4          0    -0.320000E+02   0.000000E+00
c+ 1 c+ 4  c 1  c 4          0     0.320000E+02   0.000000E+00
c+ 1 c+ 4  c 2  c 3          0     0.320000E+02   0.000000E+00
c+ 2 c+ 3  c 1  c 4          0     0.320000E+02   0.000000E+00
c+ 2 c+ 3  c 2  c 3          0     0.320000E+02   0.000000E+00
c+ 2 c+ 4  c 1  c 3          0    -0.320000E+02   0.000000E+00
c+ 2 c+ 4  c 2  c 4          0    -0.320000E+02   0.000000E+00
c+ 1 c+ 3  c 1  c 4          1    -0.640000E+02  -0.000000E+00
c+ 1 c+ 3  c 2  c 3          1    -0.640000E+02  -0.000000E+00
c+ 1 c+ 4  c 1  c 3          1    -0.640000E+02  -0.000000E+00
c+ 1 c+ 4  c 2  c 4          1    -0.640000E+02  -0.000000E+00
c+ 2 c+ 3  c 1  c 3          1    -0.640000E+02  -0.000000E+00
c+ 2 c+ 3  c 2  c 4          1    -0.640000E+02  -0.000000E+00
c+ 2 c+ 4  c 1  c 4          1    -0.640000E+02  -0.000000E+00
c+ 2 c+ 4  c 2  c 3          1    -0.640000E+02  -0.000000E+00
c+ 1 c+ 2 c+ 3  c 1  c 2  c 4          1     0.128000E+03   0.000000E+00
c+ 1 c+ 2 c+ 4  c 1  c 2  c 3          1     0.128000E+03   0.000000E+00
c+ 1 c+ 3 c+ 4  c 2  c 3  c 4          1     0.128000E+03   0.000000E+00
c+ 2 c+ 3 c+ 4  c 1  c 3  c 4          1     0.128000E+03   0.000000E+00
\end{verbatim}
The integer in the front line counts the number of terms $N_{\text{term}}$.
\texttt{c+} and \texttt{c} denote creation and annihilation operators, respectively.
The integers following \texttt{c+} and \texttt{c} are collective indexes of sites and spins.
In this example,
\texttt{1} (\texttt{3}) labels the spin up (down) on the first site
and
\texttt{2} (\texttt{4}) labels the spin up (down) on the second site.
The last two real numbers are the real and imaginary parts of
the coefficient $\beta_x$.
The integer before $\beta_x$ is $m$.

We also provide an example for $S^{k,m}$ generated by our code available at~\cite{SourceCode_sm}.
The $S^{k,m}$ for $L = 14$, $N_{f}=14$, $N_\uparrow=7$, and $l = 3$ with $U=8$ are given by
\begin{verbatim}
1.000000    8.000000  # J, U
14   14    7    7     # L, Nf, Nup, Ndn
3                     # driving order l

Gamma1
S(1,0) =    0.568273305600000E+10

Gamma2
S(2,0) =    0.148625326080000E+12
S(2,1) =    0.363694915584000E+12

Gamma3
S(3,0) =    0.533652347289600E+13
S(3,1) =    0.230526623416320E+14
S(3,2) =    0.232764745973760E+14

Gamma4
S(4,0) =    0.222388949680128E+15
S(4,1) =    0.137124173787955E+16
S(4,2) =    0.259980316272230E+16
S(4,3) =    0.148969437423206E+16

Gamma5
S(5,0) =    0.101046765965967E+17
S(5,1) =    0.816746874273792E+17
S(5,2) =    0.227024409238438E+18
S(5,3) =    0.254852329868624E+18
S(5,4) =    0.953404399508521E+17

Gamma6
S(6,0) =    0.485832012237767E+18
S(6,1) =    0.491237560046164E+19
S(6,2) =    0.182253012110310E+20
S(6,3) =    0.304507281118027E+20
S(6,4) =    0.230283831881289E+20
S(6,5) =    0.610178815685453E+19
\end{verbatim}
Here, \texttt{Gamma1}, \texttt{Gamma2}, etc. denote $\Gamma_1$, $\Gamma_2$, etc., respectively, and
only nonzero $S^{k,m}$ are shown.

\subsection*{Benchmark tests}

\begin{figure}
\includegraphics[width=\columnwidth]{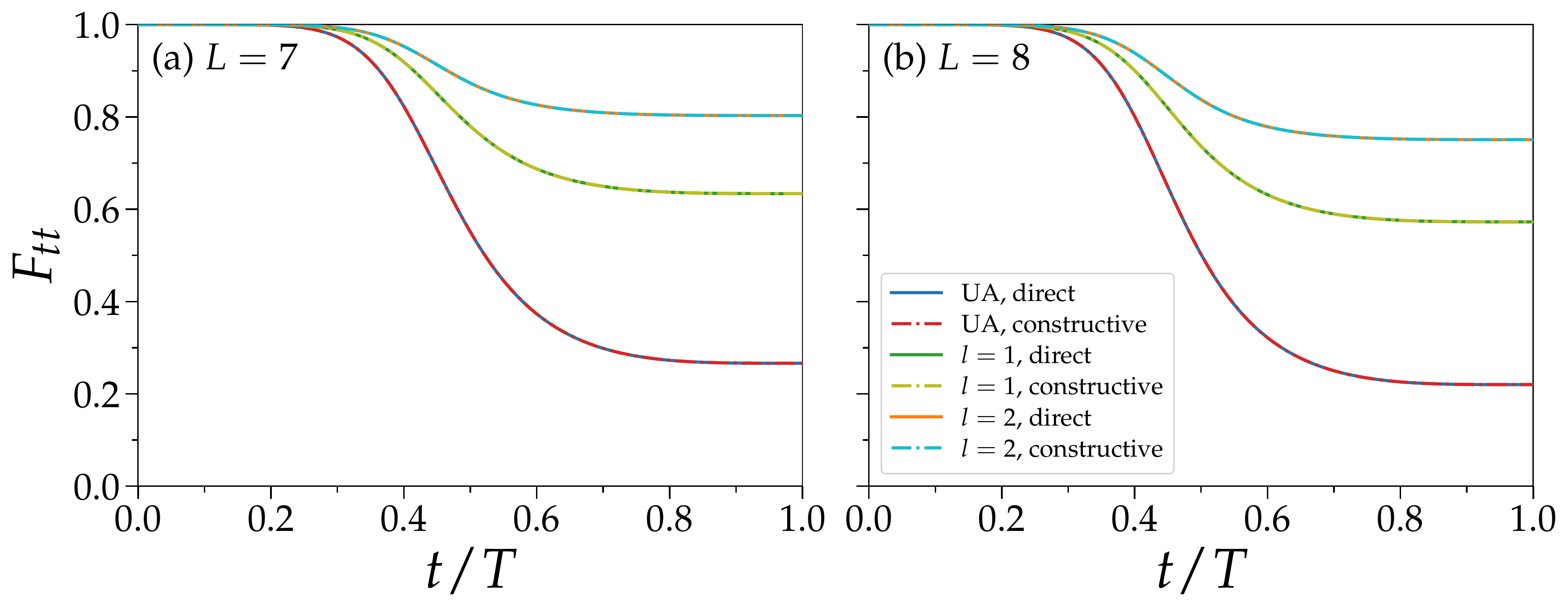}
\caption{
Comparison of the results for the time evolution of the
fidelity $F_{tt}$ obtained by the direct and constructive approaches
for the UA and CD ($l=1,2$) models with $U=8$ on
(a) $L = 7$ and (b) $L=8$ sites at half filling.
In (a), we set $N_\uparrow=N_\downarrow+1$.
The remaining parameters are the same as in Fig.~\ref{fig:fidelity} in the main text.
}
\label{fig:direct}
\end{figure}

Figure~\ref{fig:direct}
compares the results for the time evolution of the fidelity $F_{tt}$ obtained by the direct and constructive approaches
for the UA and CD ($l=1,2$) models with $U=8$ on $L = 7$ and $8$ sites at half filling. As is expected, these two results are
exactly the same within the numerical precision.

\subsection*{ $U$ dependence of $T_{\rm adi}$  }

\begin{figure}
\includegraphics[width=.8\columnwidth]{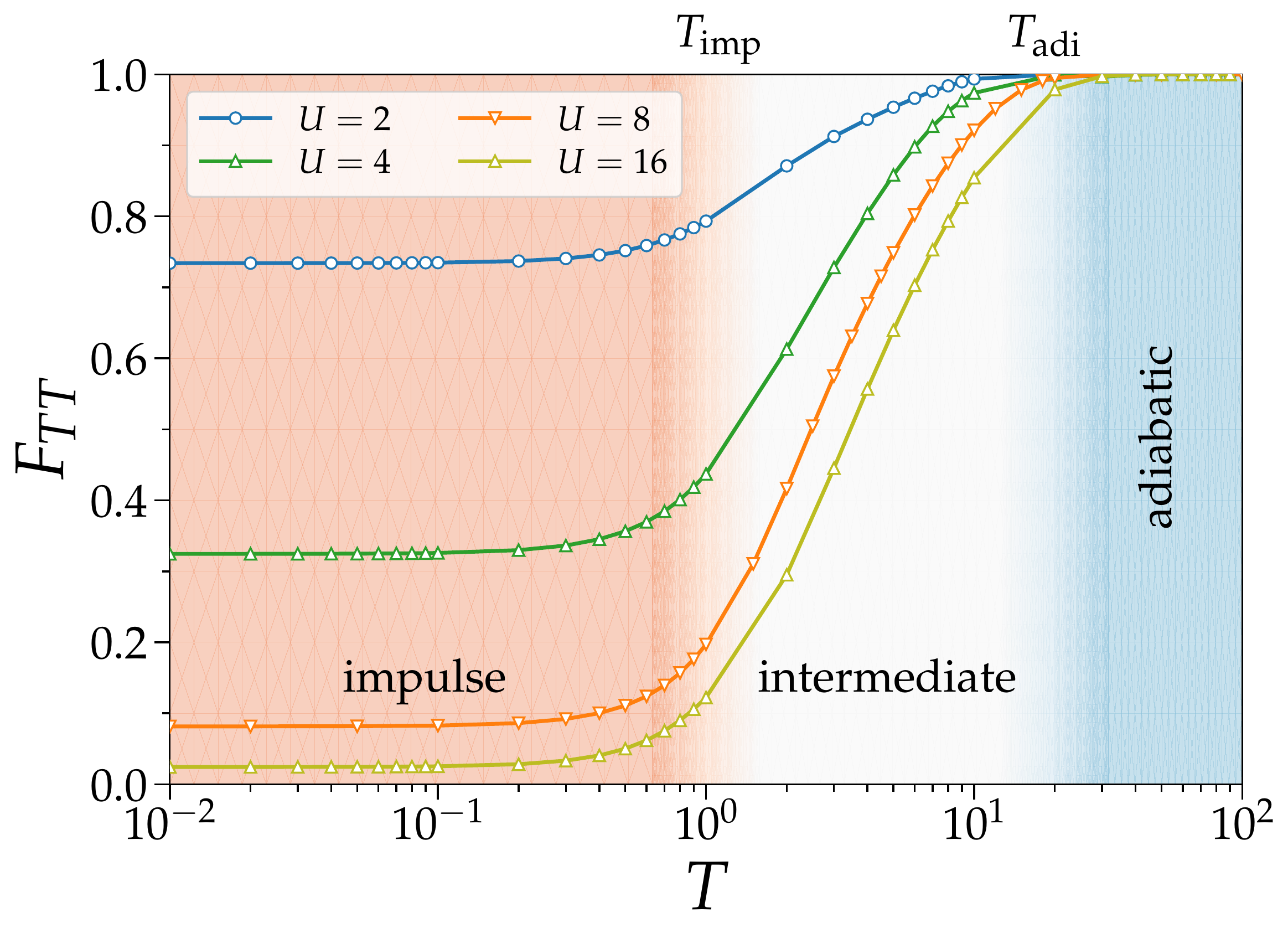}
\caption{
The final fidelity $F_{TT}$ as a function of the driving period $T$
for the UA model with $U=2,4,8$, and $16$ on $L = 12$ sites at half filling.
For various $T$, the time step is fixed to be $\Delta t=0.001$.
Notice that the results for $U=8$ are the same as those shown in Fig.~\ref{fig:tau}.
}
\label{fig:fidelity5}
\end{figure}

\begin{figure}
\includegraphics[width=.8\columnwidth]{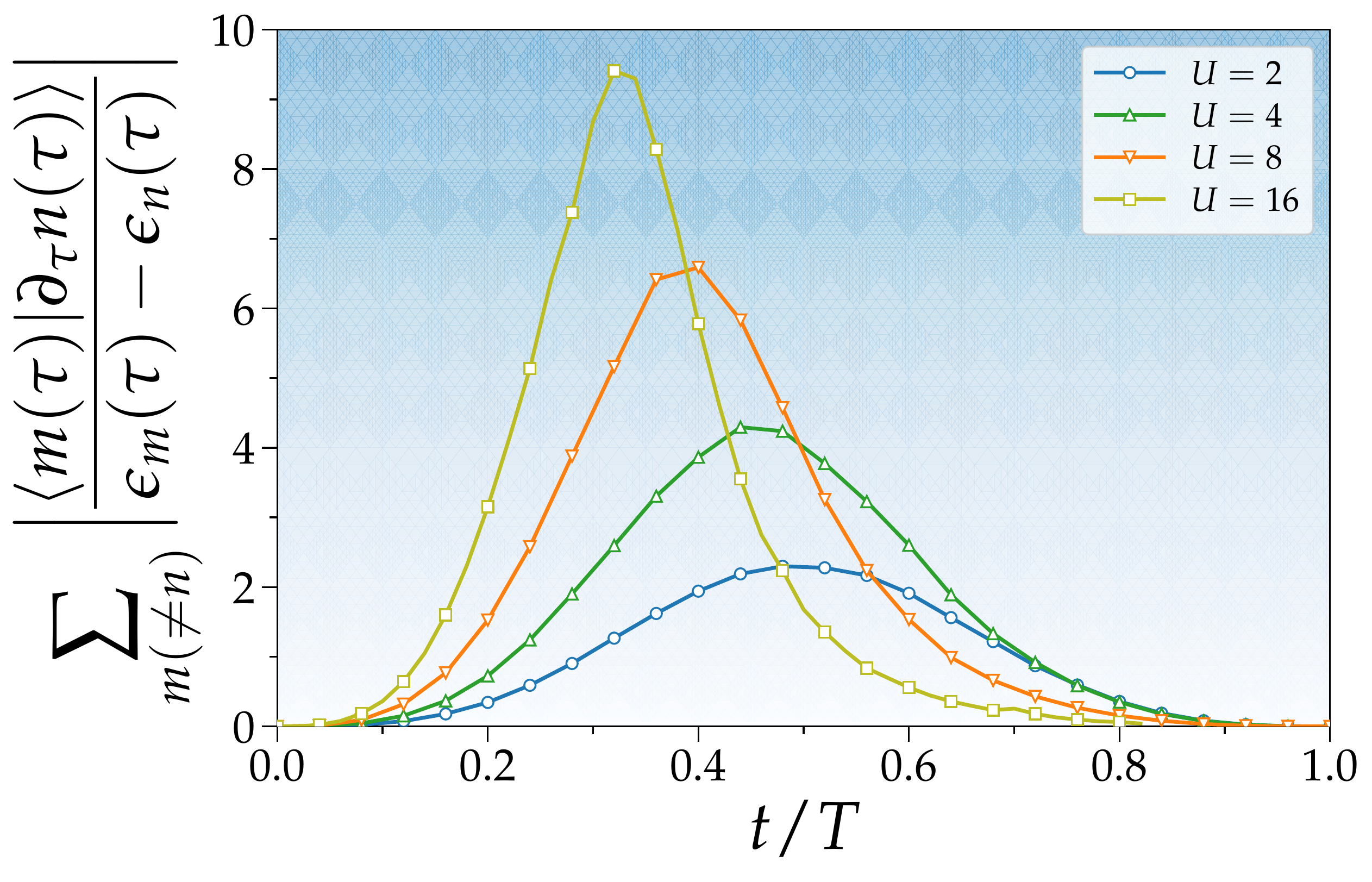}
\caption{
The time evolution of the quantity determining the adiabatic
condition given in Eq.~(\ref{eq:crit}) for the UA model
with $U = 2, 4, 8$, and $16$ on
$L = 12$ sites at half filling.
}
\label{fig:TadiU}
\end{figure}

Figure~\ref{fig:fidelity5} shows the $U$ dependence of the final fidelity $F_{TT}$ as a function of the driving period $T$
for the UA model on $L = 12$ sites at half filling. Although the crossover boundary between
the impulse and intermediate regions is insensitive to the value of the interaction strength $U$, the crossover boundary
between the intermediate and adiabatic regions depend slightly on $U$. The latter is in accordance with the $U$ dependence
of the characteristic time $T_{\rm adi}$ defined in Eq.~(\ref{eq:crit}), i.e.,
$T_{\rm adi}\sim 2, 4, 7$, and $10$
for $U=2, 4, 8$, and $16$, respectively~(see Fig.~\ref{fig:TadiU}).
Indeed, the quantity in the right hand side of Eq.~(\ref{eq:crit}) can also be written as
\begin{equation}
\label{eq:relation}
\left| \frac{ \langle m(\tau) | \partial_{\tau} n(\tau) \rangle }
{\epsilon_m(\tau) - \epsilon_n(\tau)} \right|
=
\frac{ \left|  \langle m(\tau) |\partial_\tau\hat{H}(\tau) |n(\tau) \rangle \right| }
{[\epsilon_m(\tau) - \epsilon_n(\tau)]^2}
=
\partial_\tau\lambda(\tau)
\frac{ \left| \langle m(\tau) |\hat{H}_U |n(\tau) \rangle  \right| }
{[\epsilon_m(\tau) - \epsilon_n(\tau)]^2}
=
U\partial_\tau\lambda(\tau)
\frac{ \left| \langle m(\tau) |\hat{D} |n(\tau) \rangle \right| }
{[\epsilon_m(\tau) - \epsilon_n(\tau)]^2} ,
\end{equation}
where $\partial_\tau\lambda(\tau)=T\dot{\lambda}(t)$ and
$\hat{D}=\sum_i\hat{n}_{i\uparrow}\hat{n}_{i\downarrow}$, and therefore Eq.~(\ref{eq:crit}) is now
\begin{align}
\label{eq:crit2}
T_{\rm adi} \approx
\max_{\tau \in [0, 1]}
\left[
U\partial_\tau\lambda(\tau)
\sum_{m\,(\ne n) }
\frac{ \left| \langle m(\tau) |\hat{D} |n(\tau) \rangle \right|}
{[\epsilon_m(\tau) - \epsilon_n(\tau)]^2}
\right] .
\end{align}
This implies that the characteristic time $T_{\rm adi}$ separating the intermediate and adiabatic regions is
approximately proportional to $U$, assuming that other quantities in Eq.~(\ref{eq:crit2}) do not depend strongly on $U$,
which is however not the case when $U$ is very large.

\subsection*{ Lanczos-based method for the evaluation of Eq.~(\ref{eq:crit}) }

Here, we describe a Lanczos-based method to evaluate the quantity
$\sum_{m\,(\ne n)}  \left|   \frac{ \langle m(\tau) | \partial_\tau n(\tau) \rangle } {\epsilon_m(\tau) - \epsilon_n(\tau)}   \right|  $
in the right hand side of Eq.~({\ref{eq:crit}})
through the relation in Eq.~(\ref{eq:relation}).
At each time step $t_i$,
the quantity $\sum_{m\,(\ne n)} \frac{ | \langle m(t_i)| \hat{D} | n(t_i)\rangle | }{[\epsilon_m(t_i) - \epsilon_n(t_i)]^2}  $
with $n = 0$, i.e., the instantaneous ground state of the UA model,
can be evaluated as following
(the time $t_i$ dependence is abbreviated for simplicity):

\begin{enumerate}
\item
First calculate the ground state $|0\rangle$
and its associated energy $\epsilon_0$ of the UA model $\hat{H}(t_i)$.

\item
Prepare the state $|\phi_0 \rangle = \hat{D}|0 \rangle$.

\item
Compute the normlization constant $N_0 = \sqrt{\langle \phi_0 | \phi_0 \rangle} $.

\item
Run an $M$-step Lanczos iteration for the UA model $\hat{H}(t_i)$, starting with the normalized vector
\begin{align}
| \tilde{\phi}_0  \rangle = |\phi_0\rangle / N_0  ,
\end{align}
and obtain the resultant $M\times M$ tridiagonal matrix ${\bf T}$,
which is an approximate matrix representation of the UA model
$\hat{H}(t_i)$ in the $M$-dimensional Krylov subspace.

\item
Diagonalize the tridiagonal matrix ${\bf T}$ to obtain
the eigenvalues $\{{\tilde{\epsilon}_l \}_{l = 0}^{M-1}}$ and the associated
eigenvectors $ \{ {\bf v}_{l} \}_{l = 0}^{M-1}$.

\item
The desired quantity can be evaluated approximately as
\begin{align}
\sum_{m\, (\ne 0)}  \frac{  \left| \langle m| \hat{D} | 0\rangle \right|  }{ ( \epsilon_m - \epsilon_0 )^2 }
\approx
N_0 \sum_{l = 1}^{M-1}  \frac{ \left| [{\bf v}_l ]_0 \right| }{ ( \tilde{\epsilon}_l - \epsilon_0 )^2 }.
\end{align}
Here $[{\bf v}_l ]_0$ is the first entry of the $l$th eigenvector
and
it represents the overlap between the $l$th eigenvector
${\bf v}_l $ and the normalized initial vector $|\tilde{\phi}_0\rangle$
[see, for example, Eq.~(81) in Ref.~\cite{PhysRevB.98.205114_sm}].
\end{enumerate}

\begin{figure}
\includegraphics[width=.8\columnwidth]{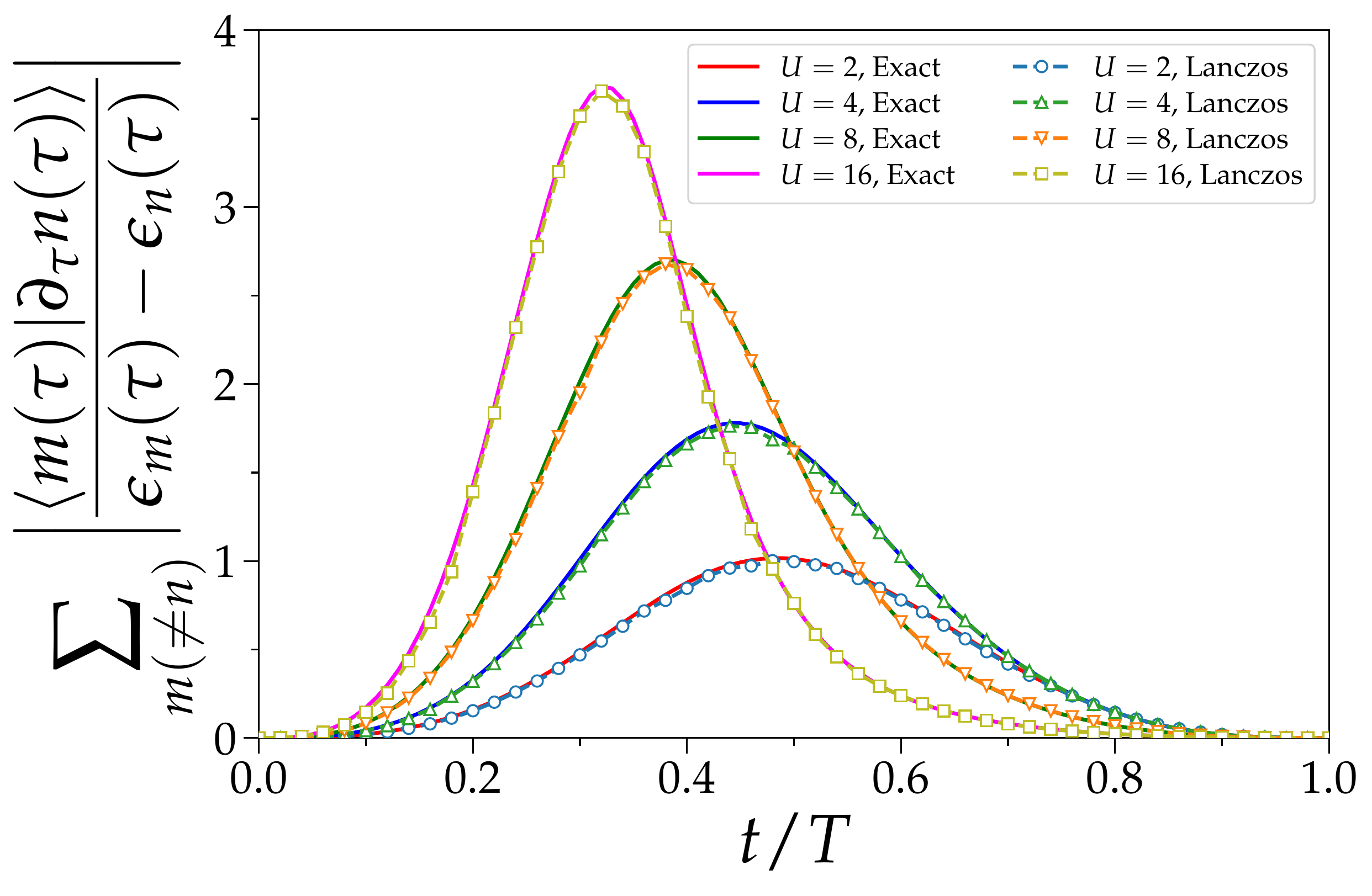}
\caption{
Comparison of the results for the time evolution of the quantity
determining the adiabatic condition given in Eq.~(\ref{eq:crit})
calculated by the Lanczos method and
the numerically exact full diagonalization method
for the UA model with $U = 2, 4, 8$, and $16$ on $L = 6$ sites at half
filling.
}
\label{fig:LvsF}
\end{figure}

We demonstrate this method for a small system with $L = 6$ sites
and compare the results with those obtained by the numerically
exact full diagonalization method in Fig.~\ref{fig:LvsF}.

\subsection*{ Fidelity of the CD model }

\begin{figure}
\includegraphics[width=.8\columnwidth]{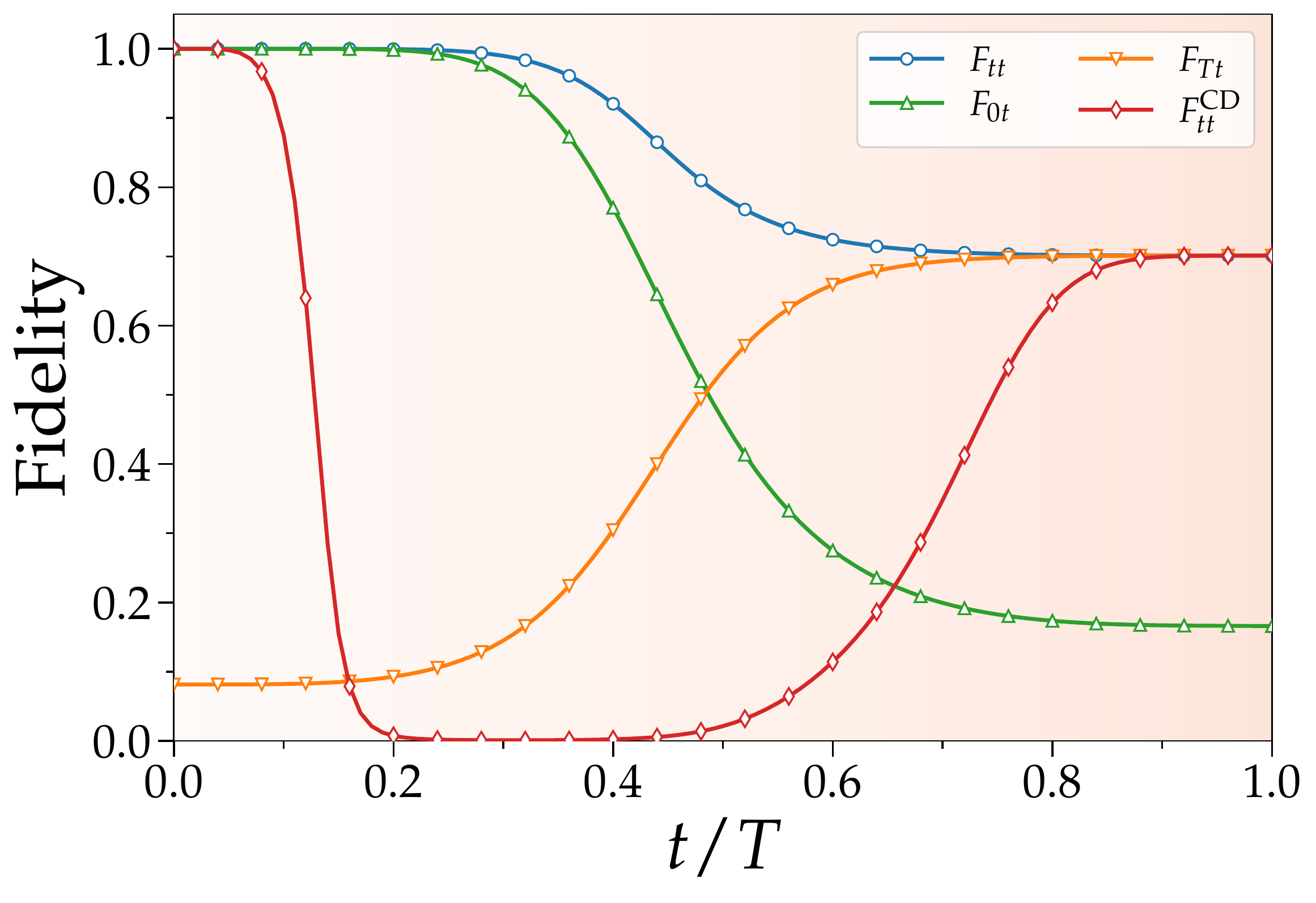}
\caption{
The time evolution of fidelity $F^{\text{CD}}_{tt}$ for the CD model with $l=3$ on the 1D chain of $L=12$ sites
at half filling.
The remaining parameters are the same as in Fig.~\ref{fig:fidelity}.
For comparison, the results for other fidelities $F_{tt}$, $F_{0t}$, and $F_{Tt}$ are also shown.
}
\label{fig:FCDtt}
\end{figure}

It is also interesting to examine the time evolution of the fidelity $F^{\text{CD}}_{tt}$ for
the CD model $\hat{H}_{\rm CD}^{(l)}$ defined by
$F^{\text{CD}}_{tt}= |\langle n^{\text{CD}}(t)| \psi(t) \rangle|^2$,
where $|n^{\text{CD}}(t)\rangle$ and $|\psi(t)\rangle$
are the instantaneous eigenstate and the time-evolved state of the CD model, respectively.
Figure~\ref{fig:FCDtt} shows the results for the CD model with $l=3$ on the 1D chain of $L = 12$ sites at
half filling.

\end{document}